\documentclass{svjour3}                     
\smartqed  
\usepackage{graphicx}
\usepackage{natbib}

\def\aap{A\&A}
\def\apj{ApJ}

\def\apjl{ApJ}
\def\mnras{MNRAS}
\def\araa{ARA\&A}
\def\aj{AJ}

\def\apjs{ApJS}

\newcommand{\mbh}{\ensuremath{M_\mathrm{BH}}}

\newcommand{\msun}{\,{\rm M_\odot}}

\newcommand{\msigma}{M_{\rm BH}-\sigma}

\newcommand{\vir}{{\rm vir}}

\newcommand{\beq}{
\begin{equation}
}
\newcommand{\eeq}{
\end{equation}
}
\newcommand{\ba}{
\begin{eqnarray}
}
\newcommand{\ea}{
\end{eqnarray}
}
\def\spose#1{\hbox to 0pt{#1\hss}}
\newcommand{\lta}{\mathrel{\spose{\lower 3pt\hbox{$\mathchar"218$}}
      \raise 2.0pt\hbox{$\mathchar"13C$}}}
\newcommand{\gta}{\mathrel{\spose{\lower 3pt\hbox{$\mathchar"218$}}
      \raise 2.0pt\hbox{$\mathchar"13E$}}}
\newcommand{\simlt}{{\mathrel{\rlap{\lower 3pt\hbox{$\sim$}}\raise 2.0pt\hbox{$<$}}}}
\newcommand{\simgt}{{\mathrel{\rlap{\lower 3pt\hbox{$\sim$}} \raise 2.0pt\hbox{$>$}}}}

\begin{document}

\title{Formation of Supermassive Black Holes
}


\author{Marta Volonteri}


\institute{M. Volonteri \at
              Astronomy Department\\
              University of Michigan\\
              500 Church Street
              Ann Arbor, MI 48109\\
              USA\\
              Tel.: +1-734-764-3440\\
              Fax: +1-734-763-6317\\
              \email{martav@umich.edu}           
}

\date{Received: date / Accepted: date}

\maketitle

\begin{abstract}
Evidence shows that massive black holes reside in most local galaxies.   Studies have also established a number of relations between the MBH mass and properties of the host galaxy such as bulge mass and velocity dispersion. These results suggest that central MBHs, while much less massive than the host ($\sim$ 0.1\%), are linked to the evolution of galactic structure.  In hierarchical cosmologies, a single big galaxy today can be traced back to the stage when it was split up in hundreds of smaller components. Did MBH seeds form with the same efficiency in small proto-galaxies, or did their formation had to await the buildup of substantial galaxies with deeper potential wells? I briefly review here some of the physical processes that are conducive to the evolution of the massive black hole population. I will discuss black hole formation processes for `seed' black holes that are likely to place at early cosmic epochs, and possible observational tests of these scenarios. 
\end{abstract}

\section{Introduction}
\label{intro}
Black holes, as physical entities, span the full range of masses, from tiny holes predicted by string theory, to monsters as massive as a dwarf galaxy. Notwithstanding the several orders of magnitude difference between the smallest and the largest black hole known, all of them can be described by only three parameters: mass, spin and charge. Astrophysical black holes are even simpler systems, as charge can be neglected. Complexities arise because of the interaction between astrophysical black holes and their environment. 

I will focus here on the formation and evolution of massive black holes (MBHs), in high-redshift galaxies, and their symbiotic evolution with their hosts.  MBHs weighing million solar masses and above have been recognized as the engines that power quasars detected at early cosmic times.  Dynamical evidence also indicates that MBHs with masses in the range $M_{BH} \sim 10^6-10^9\,\msun$ ordinarily dwell in the centers of most nearby galaxies \citep{ferrareseford}. The evidence is particularly compelling for our own galaxy, hosting a central MBH with mass $\simeq 4\times10^6\,M_\odot$   \citep[e.g.,][]{Schodel2003,Ghez2005}. MBHs populate galaxy centers today, and shone as quasars in the past; the quiescent MBHs that we detect now in nearby bulges would be the dormant remnants of this fiery past.   Dynamical estimates indicate that, across a wide range, the central black hole mass is about 0.1\% of the spheroidal component of the host galaxy \citep{Magorrian1998,MarconiHunt2003,Haring2004}.  A tight correlation is also observed between the MBH mass and the stellar velocity dispersion of the hot stellar  component \citep{fm00,Gebhardt2000,Tremaineetal2002,Gultekin2009}.  The surprisingly clear correlations between MBH masses and the properties of their host galaxies suggest a single mechanism for assembling MBHs and forming galaxies. The evidence therefore favors a common root, a co-evolution, between galaxies and MBHs.   These correlations may well extend down to the smallest masses.  For example, the dwarf Seyfert~1 galaxy POX  52 is thought to contain a MBH of mass $M_{BH} \sim 10^5\,M_\odot$ \citep{barthetal2004}. 

At the other end, however, some powerful quasars have already been detected at $z>6$, corresponding to a time less than a tenth of the age of the Universe, roughly one billion years after the Big Bang. Follow-up observations confirmed that at least some of 
these quasars are powered by super-massive black holes  with masses $\simeq 10^9\, M_\odot$ \citep{Barthetal2003,Willottetal2005}, probably residing in the centers of substantial galaxies.   However, these exceptionally bright quasars are most likely just the tip of the iceberg: rare objects-- on the tail of the mass distribution. This implies larger numbers of less exceptional objects, and that MBHs existed in large numbers during the Dark Ages, before or around the time when the first stars formed. 

\begin{figure}
\includegraphics[width=\textwidth]{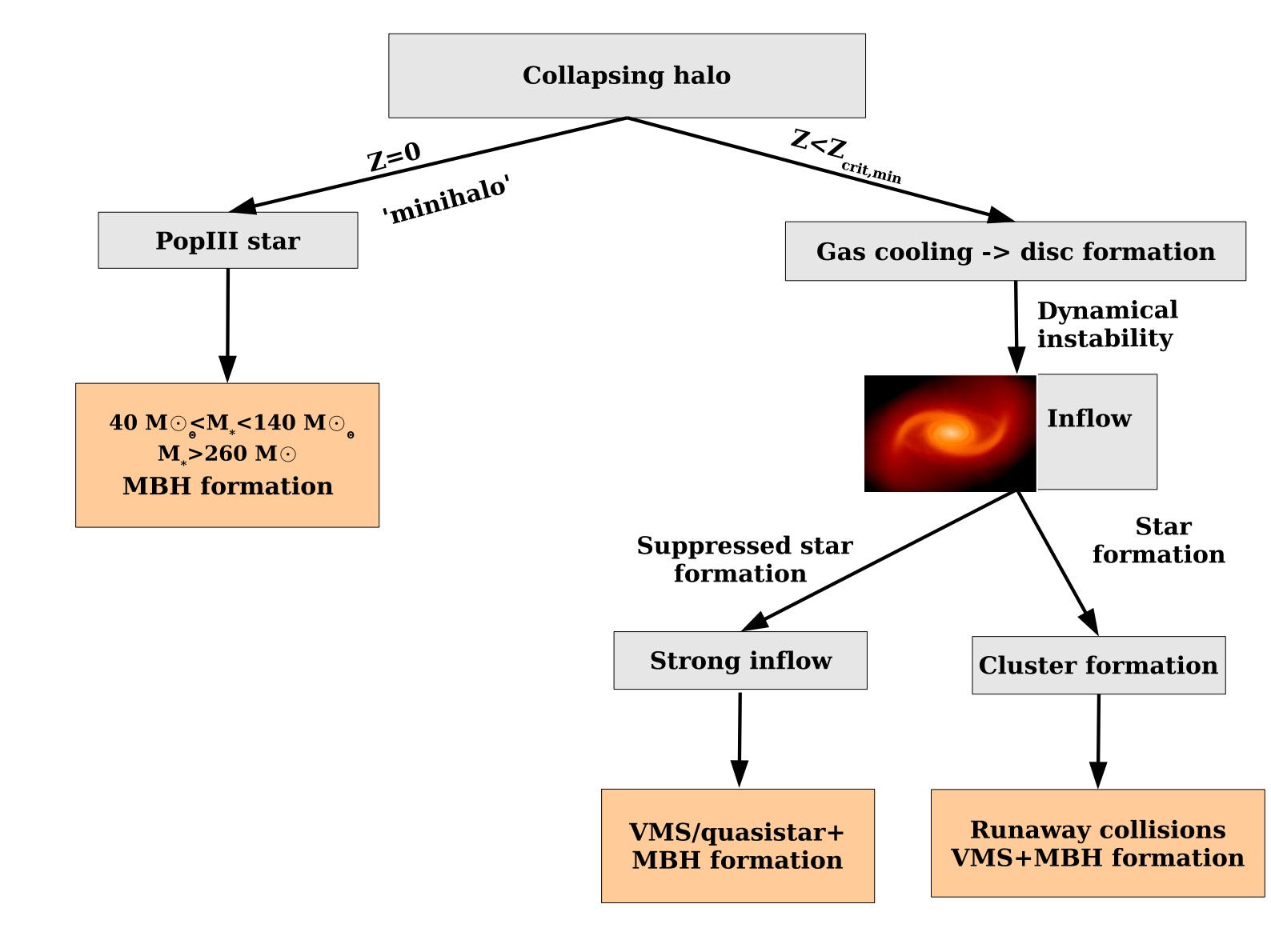}
\caption{Scheme of possible pathways to MBH formation in high-redshift galaxies. Artwork by B. Devecchi.}
\label{fig:scheme}
\end{figure}

We are therefore left with the task of explaining the presence of  MBHs when the Universe is less than {\rm 1 Gyr} old, 
and of much smaller MBHs lurking in {\rm 13 Gyr} old galaxies. The outstanding questions concern  how and when ``seed'' MBHs formed, the frequency of MBHs in galaxies, and how efficiently MBH seeds grew in mass during the first few billion years of their lives. The formation of MBHs is indeed far less understood than that of their light, stellar mass, counterparts, end-points of stellar evolution for stars more massive than $\simeq 20\,\msun$. The ``flow chart" presented by \cite{Rees1978} still stands as a guideline for the possible paths leading to formation of massive MBH seeds in the center of galactic structures. 
In the following I will review the main physical processes thought to influence MBH formation. I will mostly focus on astrophysical processes that happen in galaxies, and I will consider three possibilities: (i) that MBHs are the remnants of the first generation (PopIII) stars, (ii) that MBHs formation is triggered by gas-dynamical instabilities, (iii) that MBH seeds are formed via stellar-dynamical processes. A simplified scheme that describes the possible routes to MBH formation in high-redshift galaxies is shown in Figure~\ref{fig:scheme}.
I will also briefly mention the possibility that MBHs are related to inflationary primordial black holes.

\section{Massive black hole formation}
\label{sec:1}
\subsection{Cosmological background}
\label{ssec:1.0}
The observation of luminous quasars at $z\approx 6$ \citep[e.g,][]{Fanetal2001a} implies that the first MBHs must have formed at very early times.
The luminosities of these quasars, well in excess of $10^{47}$ erg s$^{-1}$, imply MBHs with  masses $\sim 10^9 M_\odot$ already in place when the Universe is only  1 Gyr old.   The accretion of mass at the Eddington rate causes a black hole mass to increase 
in time as
\beq 
M(t)=M(0)\,\exp\left(\frac{1-\epsilon}{\epsilon}\frac{t}{t_{\rm Edd}}\right),
\eeq 
where $t_{\rm Edd}=0.45\,{\rm Gyr}$ and $\epsilon$ is the radiative efficiency. For a `standard' radiative efficiency $\epsilon\approx0.1$, and a seed mass  $M(0)=10^2-10^5 M_\odot$, it takes at least 0.5 Gyr to grow  to $\simeq 10^9 M_\odot$. This brings us back in time to an epoch when the Universe was very young and galaxies in their infancy. 

Current theories of structure formation favour their growth from gravitational collapse of
small perturbations in a quasi-homogeneus Universe, dominated by Cold Dark Matter (CDM). 
In the framework of CDM models the collapse of structures proceeds { \it bottom-up} on larger and larger scales,
giving rise to a hierarchy of smaller structures that are incorporated into larger ones at later
times \citep{White1978}.  In the following I briefly summarize the basic points necessary to describe the cosmic landscape where MBH formation takes place.  For a detailed description of the cosmic structure formation physics, I refer the reader to \cite{Peacock1999}. 

The evolution of perturbations up to recombination and on large scales, much larger than the 
typical correlation length of the matter field, is relatively easy to follow. Such 
perturbations remain small all the time, and it is a reasonably good approximation to describe 
them as if they were in linear regime. After recombination, fluctuations continue 
growing, until they turn non-linear. At this stage structure formation takes place: collapsed
clumps form, which eventually reach an equilibrium configuration at virialization. 

By means of the virial theorem and conservation of energy the properties of virialized halos can be expressed 
as a function of the virial overdensity with respect to the background density,  $\delta_{\rm vir} \approx 178$ in an Einstein-de Sitter Universe. 
Detailed calculations for different cosmologies (e.g., $\Lambda$CDM) can be found in \cite{Lacey1993} and \cite{Eke1996}. A halo at redshift $z$ is therefore uniquely characterized by a virial radius  $r_\vir$, defined as the radius of the sphere encompassing a mean mass overdensity $\delta_\vir$. From the virial theorem the virial mass, $M_{\rm vir}$ can be calculated straightforwardly, along with the circular velocity, $V_c=\sqrt{GM_{\rm vir}/r_\vir}$ and virial temperature $T_\vir=\mu\, m_p V^2_c/(2 k_B)$ , where $\mu$ is the mean molecular weight, $m_p$ is the proton mass and $k_B$ is the Boltzmann constant. 

It is within these dark matter halos growing out of small primordial density 
fluctuations by gravitational instabilities that galaxies form out of baryons. The cosmological Jeans mass, $M_J$, represents the minimum
mass of an overdense region where pressure gradients are sufficiently weak and the 
gravitational collapse of the baryon component can proceed.  The first bound systems with $M>M_J$ capture baryons and during subsequent collapse
and virialization of the dark matter the baryons are shock heated to the virial 
temperature. The requirement for a dark matter + baryon system to become a galaxy is that star formation takes places. 
Star formation can occur if  cooling allows  the baryonic clouds to dissipate their kinetic energy, continue to collapse and fragment.

For low-mass objects, the smaller they are, the less efficiently they dissipate energy and cool. Thus a detailed treatment of gas-dynamical processes will predict a
characteristic mass scale such that more massive objects can cool rapidly, whereas smaller  lumps will merely remain pressure-supported and do not form luminous objects \citep{Tegmark1997}. The first collapsing halos have virial temperatures smaller than the $10^4$ K temperature at which cooling from electronic excitation of atomic hydrogen becomes effective. In order for their gas to cool down and form the first stars, these halos must rely on the less effective $H_2$ cooling. Halos with $T_{\rm vir} < 10^4$ K cooling by $H_2$ are typically referred to as `minihalos'.



\subsection{Population III remnants}
\label{ssec:1a}

One of the most popular scenarios for MBH formation associates MBH seeds with the remnants of the first generation of stars, formed out of zero metallicity gas.  The first stars are expected to form in minihalos, $M_{\rm min}\approx  10^6\,\msun$, above the cosmological Jeans mass collapsing at $z\sim 20-50$ from the highest peaks of the primordial density field, where cooling is possible by means of molecular hydrogen \citep[$T_{\rm vir} > 10^3$ K;][]{Tegmark1997}. Atomic hydrogen cooling has to await the condensation of larger halos, with $T_{\rm vir} \gta 10^4$K and mass $\approx  10^8\,\msun$. Above $M_{\rm min}$ the H$_2$  cooling time is shorter than the Hubble time at virialization, the gas in the  central halo regions becomes self-gravitating, and stars can form.  

The study of fragmentation of gas in primordial conditions \citep{palla2002} is characterized by some physical simplifications, as compared to present-day star formation: in the absence of metals and dust the \emph{only} coolant is hydrogen  (atomic and molecular), magnetic fields were likely to be dynamically negligible,  and the chemistry and heating of the primordial gas was not complicated by the presence  of a radiation background (excepting the cosmic microwave background).  

Simulations of the collapse of primordial  molecular clouds \citep{bromm1999,bromm2002,abel2000,Yoshida2006,gao2006}  suggest that the first generation of stars contained many `very massive stars' (VMSs) with $m_\star>100\,\msun$ \citep{CBA84}.  This is because of the slow subsonic contraction -- a regime set up by the main gas coolant, molecular hydrogen -- further fragmentation into sub-components is not seen \citep[although it is not clear if this is  a numerical effect,  rather than due to the gas physics, see][]{Glover2008a}. 

Moreover, the different conditions of temperature and density of the  collapsing cloud result in a mass accretion rate over the hydrostatic protostellar core $\sim 10^3$ times larger than what observed in  local forming stars, suggesting that Pop III stars were indeed very massive \citep{Omukai1998,Ripamonti2002, Tan2004}. If the first stars retain their high mass until death, they will collapse after a short ($\approx$ Myrs) life-time. The final fate depends on the exact mass of the star. Figure \ref{figfwh} illustrates the fate of primordial stars as a function of their initial mass. 

Between 25 and 140 $\msun$, low-metallicity stars are predicted to form black holes directly. The mass of the remnant is about half the star's mass \citep[$\sim10\msun$ in a 25 $\msun$ star and $\sim40\msun$ in a 100 $\msun$ star;][]{Zhang2008}. It is likely, however, that if the forming black hole is too light, it will not be dynamically stable within the center of its host, once stars populate a galaxy.  A light black hole might wander within its host, dynamically interacting with stars of similar mass, rather than settling at the center of the galaxy's potential well. 

Between approximately 140 and 260 $\msun$ lies the domain of pair instability supernovae. After central helium burning, stars have high
enough central entropy that they enter a temperature and density regime in which electron/positron pairs are created in abundance, converting
internal gas energy into rest mass of the pairs without contributing much to the pressure 
\citep{Barkat1967, Bond1984}. When this instability is encountered, the star contracts rapidly until implosive oxygen and silicon burning, depending on the
mass of the star, produce enough energy to reverse the collapse. These objects are completely disrupted by nuclear-powered explosions. 
The stellar core implodes to a certain maximum temperature that depends on its mass, burns fuel explosively, and explodes, leaving no remnants 
\citep{Kudritzki2000,fryer2001}.

In still more massive stars (over 260 $\msun$ on the main sequence), a new phenomenon occurs. A sufficiently large fraction of the center
of the star becomes so hot that the photodisintegration instability is encountered before explosive burning reverses the implosion \citep{Bond1984} . This uses up all the energy released by previous burning stages and, instead of producing an explosion, accelerates the collapse.  The nuclear energy released by pairs when the star collapses is not sufficient to reverse the implosion before the onset of the photodisintegration instability, and the star becomes a black hole \citep{Bond1984,Woosley1986}, sweeping all heavy-element production inside. A massive black hole, containing at least half of the initial stellar mass, is born inside the star \citep{fryer2001}. 

If VMSs form above 260 $\msun$, after $\sim$ 2 Myr they would therefore collapse to MBHs with masses intermediate between those of the stellar and supermassive variety. It has been suggested by \cite{MadauRees2001} that a numerous population of MBHs may have been the endproduct of the first episode of  pregalactic star formation; since they form in  density high peaks, relic MBHs with mass $\gta 150\,\msun$ would be predicted to cluster in the cores of more massive halos formed by subsequent mergers.


Although this path to MBH formation seems very natural, large uncertainties exist on the final mass of PopIII stars. Despite the large number of studies performed, the initial mass function of PopIII is still not strongly constrained. As mentioned, it is not clear if the very first stars formed in singles or multiples per halo, with the latter option leading to a less top-heavy mass function \citep{Glover2008,Stacy2009,Turk2009}. And, if single objects are formed, simulations of the initial phase of formation of these objects only show that massive $\sim \,10^3\,\msun$ clumps of gas can collapse leading to the formation of a very dense, optically thick core of $\approx 0.01\,\msun$. Gas in the envelope accretes into the core, increasing the mass of the protostar. The characteristic final mass at the end of the accretion process can be much less than the initial clump mass as feedback effects can act, strongly reducing the ability of the core to accrete material \citep{McKee2008}. In addition, the characteristic mass of the star can depend on additional external factors as the presence of external UV radiation and the temperature of the external cosmic microwave background floor \citep[see the discussion in][]{Trenti2009}. We do not know if PopIII stars are indeed very massive, and in particular if they are above the threshold ($\simeq 260\msun$) for MBH formation. 

It has also been proposed that  dark matter, in the form of weakly interacting massive particles (WIMPs, e.g. neutralinos) can influence the formation of the first stars \citep{Ripamonti2007,Iocco2008}. If dark matter halos have steep density profiles \citep[e.g., NFW profile][]{NFW1997}, during the formation of the first stars the baryonic infall compresses the dark matter further (e.g., adiabatic contraction).  The central dark matter densities can be high enough that WIMP annihilation can provide an extra heat source. The energy  released by the annihilation of WIMPs in the stellar core overcomes cooling processes and halts stellar collapse. The resulting object, a `Dark Star', is supported by dark matter annihilation rather than nuclear fusion \citep{Spolyar2008,Freese2008}. Dark stars are predicted to be $\sim 500-1000 \msun$, with large luminosity ($10^6-10^7 L_\odot$), and low surface temperature ($ < $10,000 K).  Such massive systems would collapse into MBHs, providing alternative MBH seeds.


\subsection{Gas-dynamical processes}
\label{ssec:1b}
Another family of models for MBH formation relies on the collapse of supermassive objects formed directly out of dense gas \citep{haehnelt1993,LoebRasio1994, Eisenstein1995,BrommLoeb2003,Koushiappas2004, BVR2006,LN2006}. The physical conditions (density, gas content) in the inner regions of mainly gaseous proto-galaxies make these locii natural candidates, because the very first proto-galaxies were by definition metal-free, or at the very least very metal-poor. Enriched halos have a more efficient cooling, which in turn favors fragmentation and star formation over the efficient collection of gas conducive to MBH formation. It has been suggested that efficient gas collapse probably occurs only in massive halos with virial temperatures $T_{\rm vir} \gta 10^4$K under metal-free conditions where the formation of H$_2$ is inhibited \citep{BrommLoeb2003}, or for gas enriched below the critical metallicity threshold for fragmentation \citep{santoro}. For these systems the tenuous gas cools down by atomic hydrogen only until it reaches $T_{\rm gas}\sim 4000$ K.  At this point the cooling function of the atomic hydrogen drops by a few orders of magnitude, and contraction proceeds nearly adiabatically. 

Suppressing molecular-hydrogen formation in metal--free galaxies requires, broadly speaking, the presence of  dissociating UV radiation. The critical UV fluxes required to suppress H$_2$ in massive halos ($T_{\rm vir} \gta 10^4$K) are indeed high compared to the expected level of the cosmic UV background at the redshifts of interest. \cite{Dijkstra2008} suggest that halos prone to H$_2$ suppression are the small subset of all halos (estimated at $\sim10^{-6}$) that sample the bright--end tail of the fluctuating cosmic UV background, due to the presence of a close luminous neighbour. \cite{Spaans2006} suggest instead an alternative route: for isothermally collapsing gas at $T_{\rm vir} \gta 10^4$K line trapping of  Lyman $\alpha$ photons causes the equation of state to stiffen, making fragmentation harder, in systems with metallicity below $\sim 10^{-4}$ solar. H$_2$ is naturally destroyed in these systems, because of the high gas temperature caused by Lyman $\alpha$ trapping.  
Finally, it has been recently suggested that highly turbulent systems are also likely to experience a limited amount of fragmentation, suggesting that efficient gas collapse could proceed also in metal-enriched galaxies at later cosmic epochs \citep{Begelman2009}. 

In such halos where fragmentation is suppressed, and cooling proceeds gradually,  the gaseous component can cool and contract until rotational support halts the collapse.   In the most common situations, rotational support can halt the collapse before densities required for MBH formation are reached. Halos, and their baryonic cores, possess  angular momentum, $J$, believed to be acquired by tidal torques due to interactions with neighboring halos. This can be quantified through the so-called spin parameter, which represents the degree of rotational support available in a gravitational system: $\lambda_{\rm spin} \equiv J |E|^{1/2}/G M_h^{5/2}$, where $E$ and $M_h$ are the total energy and mass of the halo.  

Let $f_{\rm gas}$ be the gas fraction of a proto-galaxy mass, and $f_d$ the fraction of the gas that can cool. A mass $M=f_d\,f_{\rm gas}\,M_h$ would then settle into a rotationally supported disc \citep{MoMaoWhite1998,OhHaiman2002} with a scale radius $\simeq  \lambda_{\rm spin} r_{\rm vir}$, where $r_{\rm vir}$ is the virial radius of the proto--galaxy. Spin parameters found in numerical simulations are distributed log-normally in $\lambda_{\rm spin}$, with mean  $\bar \lambda_{\rm spin}=0.04$ and standard deviation $\sigma_\lambda=0.5$ \citep{Bullock2001,VandenBosch2002}. In a typical high-redshift galaxy with $T_{\rm vir} \gta 10^4$K ($M_h\approx  10^8\,\msun$; $r_{\rm vir} \approx 500$ pc) the tidally induced angular momentum would therefore be enough to provide centrifugal support at a distance $\simeq 20$ pc from the center, and halt collapse, ultimately leading to the formation of a disk. Additional mechanisms inducing transport of angular momentum are needed to further condense the gas until conditions fostering MBH formation are achieved. 

\cite{Eisenstein1995} and \cite{Koushiappas2004} investigated the formation of black holes from low angular momentum material, either in halos with extremely low angular momentum, or by considering only the low angular momentum tail of material in halos with efficient gas cooling.  But even in these models, as in all the others, substantial angular momentum transport is required in order for the gas to form a central massive object, which ultimately collapses  as a result of the post-Newtonian gravitational instability.  

An appealing route to efficient angular momentum shedding is by global dynamical instabilities, such as the ``bars-within-bars" mechanism, that relies on global gravitational instability and dynamical infall \citep{Shlosman1989, BVR2006}.  Self-gravitating gas clouds become bar-unstable when the level of rotational support surpasses a certain threshold. A bar can transport angular momentum outward on a dynamical timescale via gravitational and hydrodynamical torques, allowing the radius to shrink.  Provided that the gas is able to cool, this shrinkage leads to even greater instability, on shorter timescales, and the process cascades.  This mechanism is a very attractive candidate for collecting gas in the centers of halos, because it works on a dynamical time and can operate over many decades of radius. 

\begin{figure} 
\includegraphics[width=0.75\columnwidth]{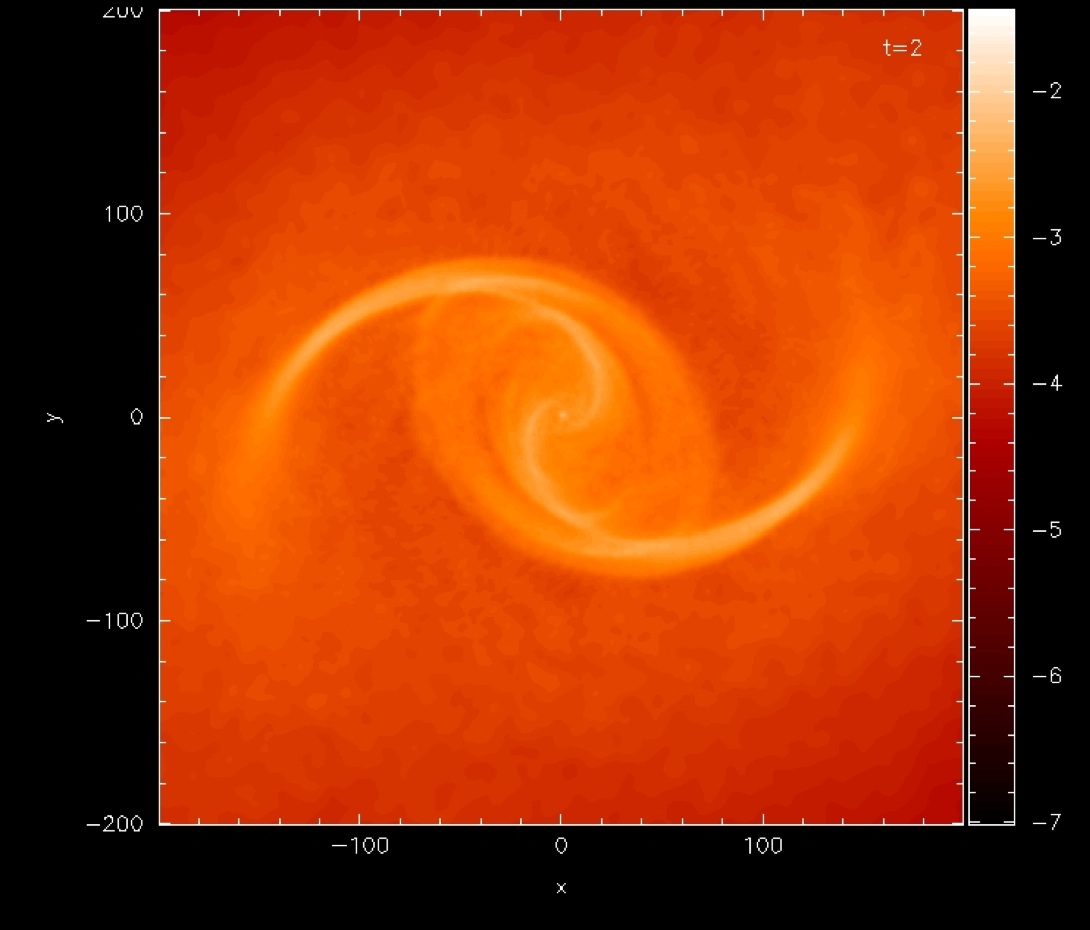}
\includegraphics[width=0.75\columnwidth]{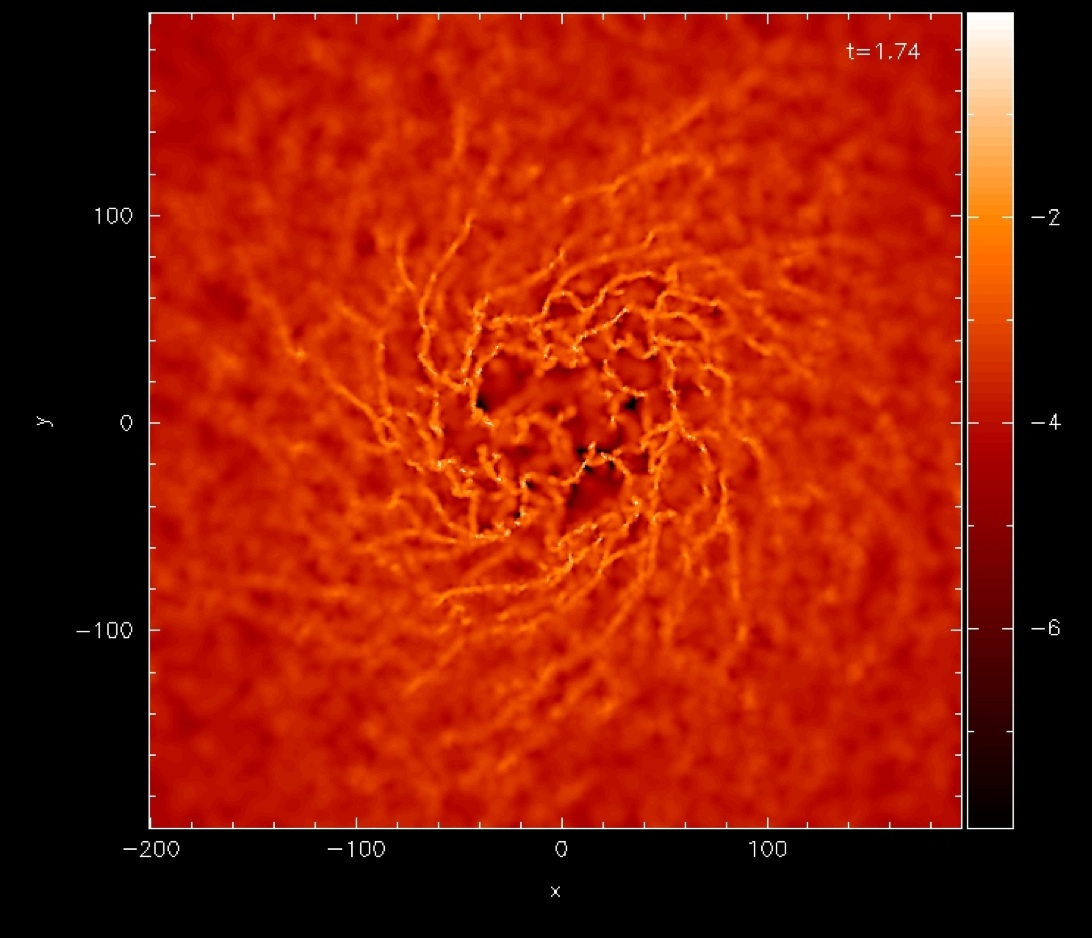}
  \caption{Color-coded density maps for two exponential gaseous discs, each embedded in a dark matter halo, soon after the discs reach the instability regime. Upper panel: in this case, fragmentation is suppressed, spiral waves develop in the disc, funnelling gas in the central few parsecs. Lower panel:  strong cooling allows fragmentation in the disc to set in. }
   \label{fig:berni}
\end{figure}

It has been also proposed that gas accumulation in the central regions of protogalaxies can be described by local, rather than global, instabilities. During the assembly of a galaxy disc, the disc can become self-gravitating. As soon as the disc becomes massive enough to be marginally stable, it will develop structures that will redistribute angular momentum and mass through the disc, preventing the surface density from becoming too large and the disc from becoming too unstable. To evaluate the stability of the disc, the Toomre stability parameter formalism can be used. The Toomre parameter
 is defined as $Q=\frac{c_{\rm s}\kappa}{\pi G \Sigma}$, where $\Sigma$ is the  surface mass density, $c_{\rm s}$ is the sound speed,  $\kappa=\sqrt{2}V_{\rm h}/R$ is the epicyclic frequency, and $V$ is the circular velocity of the disc. When $Q$ approaches a critical value, $Q_c$, of order unity the disc is subject to gravitational instabilities. If the destabilization of the system is not too violent, instabilities lead to mass infall instead of fragmentation into bound clumps and global star formation in the entire disk \citep{LN2006}.    This is the case if the inflow rate is below a critical threshold $\dot{M}_{ max}=2\alpha_{c}\frac{c^3_s}{G}$ the disk is able to sustain (where $\alpha_c\,\sim\, 0.12$ describes the viscosity) and molecular and metal cooling are not important.  Such an unstable disc develops non-axisymmetric spiral structures, which effectively redistribute angular momentum, causing mass inflow (Figure~\ref{fig:berni}, upper panel).   This process stops when the amount of mass transported to the center is enough to make the disc marginally stable. The mass that has to be accumulated in the center to make the disc stable, $M_a$, is obtained by requiring that $Q=Q_c$. This condition can be computed from the Toomre stability criterion and from the disc properties, determined from the dark matter halo mass, via $T_{\rm vir}\propto M_h^{2/3}$, and angular momentum, via the spin parameter, $\lambda_{\rm spin}$, defined above \citep{MoMaoWhite1998}:
\begin{equation}
M_a= f_{\rm d}M_{\rm halo}\left[1-\sqrt{\frac{8\lambda}{f_{\rm d}Q_{\rm c}}\left(\frac{j_{\rm d}}{f_{\rm d}}\right)\left(\frac{T_{\rm gas}}{T_{\rm vir}}\right)^{1/2}}\right]. 
\label{mbh}
\end{equation}
for $\lambda<\lambda_{\rm max}=f_{\rm d}Q_{\rm c}/8(f_{\rm d}/j_{\rm d}) (T_{\rm  vir}/T_{\rm gas})^{1/2}$. Here $\lambda_{\rm max}$ is the maximum
halo spin parameter for which the disc is gravitationally unstable,  $f_d$ is the gas fraction that participates in the infall, and $j_d$ is the fraction  of the halo angular momentum retained by the collapsing gas.  Given the mass and spin parameter of a halo, the mass that accretes to the center in order to make  the disc stable is an upper limit to the mass that can go into MBH formation.  The efficiency of the mass assembly process ceases at large halo masses, where the mass-accretion rate from the halo is above the critical threshold for fragmentation and the disc undergoes global star formation instead (Figure~\ref{fig:berni}, lower panel).     

\cite{Wise2008} and \cite{Regan2009} have performed some idealized numerical simulations of the evolution of gaseous discs in haloes with suitable characteristics.  In their simulations, the gas can then lose up to 90\% of its angular momentum due to supersonic turbulent motions, settling eventually into rotationally supported fat discs, in a very compact configuration. These fat, self-gravitating discs are marginally stable against gravitational instabilities and resemble the dense structure predicted by the analytical calculations. Additional numerical experiments, including all the relevant physics, are needed to further explore these possible MBH nurseries. 

After it has efficiently accumulated in the center, the gas made available in the central compact region can form a central massive object. The typical masses of gas collected within the central few parsecs are of order $10^4-10^6 \, \msun$. We expect that, depending on how fast and efficiently the mass accumulation proceeds, the exact outcome would differ.  The gas might form a supermassive star (SMS, mass above $\simeq 5\times 10^4 \,\msun$), which would eventually collapse and form a black hole. The evolution of a SMS depends on whether nuclear reactions are taken into account and whether the star has a fixed mass or grows via accretion during evolution. It is generally thought that SMSs of fixed mass, supported by radiation pressure, will evolve as an $n=3$ polytrope \citep{hoyle1963,Baumgarte1999,saijo2002}. Uniformly rotating SMSs cool and contract until they reach the instability point\footnote{We refer the reader to \cite{Shapiro2004} for a comprehensive description of the evolution of a SMS of fixed mass to collapse.}. 
\cite{Baumgarte1999} investigated the structure and stability of a rapidly rotating SMS in equilibrium. As long as the viscous or magnetic braking timescale for angular momentum transfer is shorter than the evolutionary timescale of a typical SMS \citep{Zeldovich1971}, the star will rotate as a rigid body, cool and contract until it reaches the mass-shedding  (where the equatorial angular velocity equals the Kepler velocity). \cite{saijo2002} investigated the collapse of a rotating SMS in a post-Newtonian approximation to investigate whether the large amount of rotation, causes a rotating SMS to form a disk, or if it can collapses directly to a black hole.  Rotation seems unable to halt the collapse and a MBH is likely to be formed. However, a post-Newtonian calculation cannot follow the collapse into the strong-field regime and cannot rigorously address the outcome of the collapse. The fate of a marginally unstable, maximally rotating SMS of arbitrary mass has been investigated numerically in full general relativity by \cite{Shibata2002}. They found that the final object is a Kerr-like black hole (spin parameter $\approx$0.75) containing 90\% of the stellar mass. The fate of an isolated SMS is therefore the formation of a MBH.

If the mass accumulation is fast, however, the outer layers of SMSs are not thermally relaxed during much of the main sequence lifetime of the star \citep{Begelman2009b}. Gas falls onto the star with increasing specific entropy as a function of time leading to entropy stratification. Such SMSs are not necessarily well-represented by $n = 3$ polytropes, but rather can have a more complex structure with a convective (polytropic) core surrounded by a convectively stable envelope that contains most of the mass. Hydrogen burning in the core starts when the mass and entropy of the star are both relatively low, and adjusts to sustain the star through its more massive stages. After exhausting its hydrogen, the core of a SMS will contract and heat up until it suffers catastrophic neutrino losses and collapses. The initial black hole, with mass of a few $\msun$, formed as a result of core-collapse subsequently grows via accretion from the bloated envelopes that result  \citep[`quasistars', an initially low-mass black hole embedded within a massive, radiation-pressure-supported envelope; see also][]{BVR2006,Begelman2008}. Over time, the black hole grows at the expense of the envelope, until finally the growing luminosity succeeds in unbinding the envelope and the seed MBH is unveiled. The key feature of this scenario is that while the black hole is embedded within the envelope, its growth is limited by the Eddington limit for the whole quasi-star (this can be understood in analogy with normal stars, where the central core produces enough energy to sustain the whole star, rather than the core only), rather than that appropriate for the black hole mass itself. Very rapid growth can then occur at early times, when the envelope mass greatly exceeds the black hole mass. 

The rate at which mass is  supplied to the black hole's sphere of influence, $\dot M_{\rm sup}$,  can be taken to be proportional to the Bondi rate evaluated at the black hole radius of influence.  The rate at which matter actually reaches the black hole is suppressed  due to the back reaction of the energy flux inside the radius of influence 
\citep{Gruzinov1998,Blandford1999,Narayan2000,Quataert2000}. In the absence of a wind that removes energy and/or angular momentum, the accretion rate is reduced to $\dot M_{\rm BH} \sim \epsilon^{-1} (c_s/c)^2 \dot M_{\rm sup}$ \citep{Blandford1999,Blandford2004}.  Comparing the accretion rate to the inflow rate onto the quasistar, the feedback energy equals the binding energy of the quasistar before the black hole mass has doubled. Thus Bondi accretion, even modified by feedback and a finite rate of angular momentum transport, should quickly bring the quasistar to the point where its evolution is driven by feedback from the black hole.  

The feedback flux does not blow apart the quasistar, since this would stop the growth of the black hole and therefore the feedback.  Instead the quasistar expands gradually, allowing the black hole accretion rate to adjust so that the feedback energy flux equals the Eddington limit for the instantaneous quasistar mass, $\dot M_{\rm BH} \sim 2 \times 10^{-2} (\epsilon/0.1)^{-1}(M_*/10^6\msun) M_\odot$ yr$^{-1}$. The feedback energy flux exceeds the Eddington limit for the black hole by a factor $M_*/M_{\rm BH}$; thus, the black hole grows at a super-Eddington rate as long as $M_* > M_{\rm BH}$.   If the quasistar mass continues to increase at a constant rate: 
\beq
\dot M_{\rm BH}\simeq 10^{-1} \epsilon \frac{M_*}{10^6\msun}\msun {\rm yr}^{-1}
\eeq
i.e., $M_{\rm BH} \propto M_*^2$.  As the growth of black hole and quasistar proceed, the photospheric temperature of the star decreases, until it reaches a floor minimum temperature ($\sim4000$ K) below which no hydrostatic solution for the convective envelope exists, causing the convective zone to release radiation at a super-Eddington rate. After a quasistar reaches this limiting temperature, it is rapidly dispersed, setting the final limit of the black hole seed
\beq
M_{\rm BH}=\frac{6\times10^4}{\epsilon^{1/2}}\left(\frac{4000K}{T_{\rm floor}}\right)^{5/2}\left(\frac{M_*}{10^6\msun}\right)^{7/8},
\eeq 
leaving behind a seed of $\sim 10^4 \msun$. The masses of the seeds predicted by different models of gas infall and SMS structure vary, but they are typically in the range $M_{BH} \sim 10^4-10^5\,M_\odot$ (Figure~\ref{fig:MF}). 

\subsection{Stellar-dynamical processes}
\label{ssec:1c}
Efficient gas collapse, leading to MBH seed formation, is mutually exclusive with star formation, as competition for the gas supply limits the mass available. If star formation proceeds in small mini-halos ($T_{\rm vir} < 10^4$K), triggered by H$_2$ cooling \citep{bromm1999,bromm2002,abel2000}, then by the time more massive halos are built-up, they will have been enriched with some traces of metals, brought in by their progenitors. Fragmentation and formation of low mass stars starts as soon as gas is polluted by metals created in the first PopIII stars \citep{omukai2008}. Formation of `normal' stars opens up a new scenario for MBH seed formation, if  stellar-dynamical rather than gas-dynamical processes are at play.  This first episode of efficient star formation can  foster the formation of very compact nuclear star clusters \citep{schneider2006,clark2008} where star collisions can lead to the formation of a VMS, possibly leaving a MBH remnant with mass in the range $\sim\,10^2-10^4\,M_{\odot}$ \citep{Devecchi2009}.

The possibility that an MBH could in principle form as a result of dynamical interactions in dense stellar systems is a long standing idea \citep{Begelman1978, ebisuzaki2001,Miller2002,PZ2002,PZ2004,freitag2006a,freitag2006b,ato2004,ato2006}.  During their lifetime collisional stellar systems evolve as a result of dynamical interactions. In an equal mass system the central cluster core initially contracts as the
system attempts to reach a state of thermal equilibrium: energy conservation leads to a decrease in the core radius as evaporation of
the less bound stars proceeds. As a result the central density increases and the central relaxation time decreases. The core then decouples thermally from the outer region of the cluster. Core collapse speeds up as it is driven by energy transfer from the central denser region \citep{Spitzer1987}.
This phenomenon is greatly enhanced in multi-mass systems like realistic star clusters. In this case the gravothermal collapse
happens on a shorter timescale as dynamical friction causes the more massive stars of mass $m$, to segregate in the centre on a time-scale
$\simeq(\langle m\rangle/m)\,t_{h}$ (where $t_{h}$ is the relaxation timescale at the half mass radius, R$_{\rm h}$, and $\langle m\rangle$ is the mean mass of a star in the cluster). If mass segregation sets in before the more massive stars evolve out of the main sequence ($\sim$ 3 Myr), then a sub-system decoupled from the rest of the cluster can form, where star-star collisions can take place in a runaway fashion that ultimately leads to the growth of a VMS \citep{PZ1999} over a short timescale:
\begin{equation}
t_{cc}\simeq
3{\rm Myr}\,\left(R_{\rm h}\over 1{\rm pc}\right)
^{3/2}\left(M_{cl} \over 5\times10^5\,M_\odot \right)^{1/2}\times 
 \left(  10\,M_\odot  \over  \langle m \rangle \right).
\label{eq:tcc}
\end{equation}

 \cite{Gaburov2009} investigate the process of runaway collisions directly with hydrodynamical simulations  and show that during  collisions large mass losses are likely, especially at metallicities larger than 10$^{-3}$ solar.  Metal-enriched VMSs are therefore expected to end their lives as objects less massive than $\sim 150 M_\odot$, collapsing into MBHs with mass $\sim$ few $M_\odot$ or exploding as pair-instability supernovae.

\begin{figure} 
\includegraphics[width=\columnwidth]{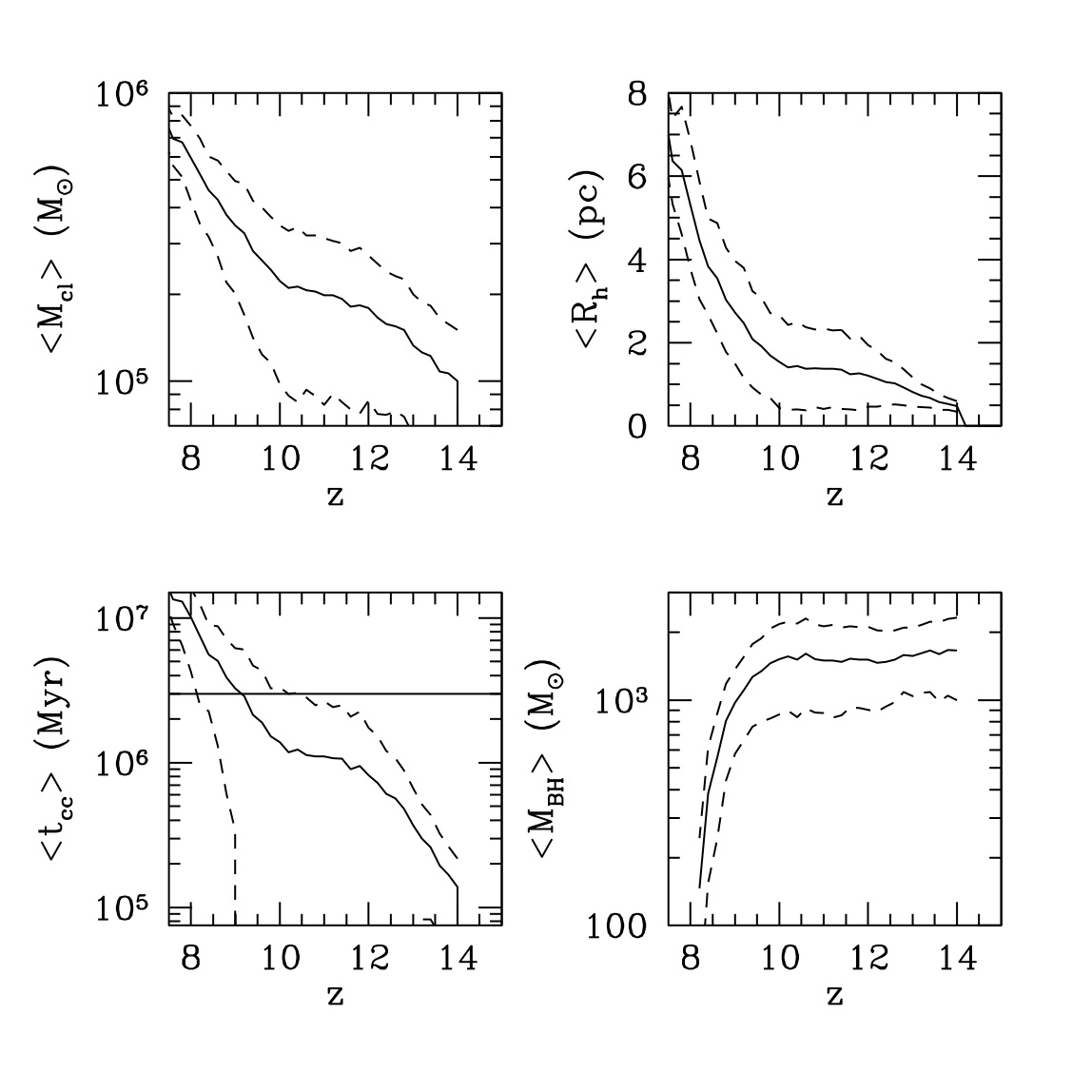}
\caption{Mean cluster masses (upper left panel), radii (upper right),
  core collapse timescales (lower left) and BH masses (lower right) as
  a function of redshift (from Devecchi \& Volonteri 2008). The orizontal line in the lower
  left panel marks the critical core collapse timescale for VMS
  formation. The dashed curves mark the 1--$\sigma$ scatter. }
\label{fig:PC} 
\end{figure}

The growth of a VMS should be much more efficient at low metallicity.  Low metallicity can modify the picture in different ways.  First,  at sub-solar (but not primordial) metallicity, all stars with masses $\gta40M_\odot$ are thought to collapse directly into a black hole without exploding as supernovae \citep{heger2003}.  Second, the mass loss due to winds is much more reduced in metal-poor stars, which greatly helps in increasing the mass of the final remnant.

\cite{Devecchi2009} investigate the formation of MBHs, remnants of VMS formed via stellar collisions in the very first stellar clusters at early cosmic times.  The main features of their model are as follows: they considered halos with virial temperatures $T_{vir}\simgt10^4$ K  after  the first episode of star formation, hence with a low, but non-zero, metallicity gas content.  This set of assumptions  ensures that (i) atomic hydrogen cooling can contribute to the gas  cooling process, (ii) a UV field has been created by the first stars,  and (iii) the gas inside the halo has been mildly polluted by the  first metals.  The second condition implies that at low density H$_2$  is dissociated and does not contribute to cooling. The third condition ensures that gas can fragment and form low-mass stars only if the gas density is above a certain threshold, $n_{crit,Z}$ \citep[which depends on the metallicity, see][]{santoro}: this causes only the highest density regions of a proto-galactic disc (see section 2.3) to be prone to star formation. 

In Toomre-unstable proto-galactic discs, such as those described in the previous section, instabilities lead to mass infall instead of fragmentation into bound clumps and global star formation in the entire disk. The gas inflow increases the central density, and within a certain, compact, region  $n>n_{crit,Z}$. Here star formation ensues and a dense star cluster is formed. At metallicities $\sim 10^{-4}-10^{-5}$ solar, the typical star cluster masses are of order $10^5\msun$ and the typical half mass radii $\sim 1$ pc.  Most star cluster therefore go into core collapse in $\simlt$ 3 Myr, and runaway collisions of stars form a VMS \citep[see Section 2.3 and][]{Shapiro2004}, leading to a MBH remnant. Figure~\ref{fig:PC}  summarizes the main characteristics of the cluster and MBH populations. 

As the  metallicity of the Universe increases, the critical density for  fragmentation decreases and stars start to form in the entire  protogalactic disk.  As gas is consumed by star formation, the inflow of gas is no longer efficient, more extended clusters form, and  the core collapse timescale increases (cfr. Equation~\ref{eq:tcc}). \cite{Devecchi2009}  find that typically a fraction $\sim 0.05$ of protogalaxies at $z\sim10-20$  form black hole seeds, with masses $\sim 1000-2000 \msun$, leading to  a mass density in seeds of a few $\simeq10^2\msun/{\rm   Mpc}^{-3}$.  Most of the assumptions in \cite{Devecchi2009} have been conservative, but still the population of seeds is comparable to the case of Population III star remnants discussed, for instance, in \cite{VHM}. The fraction of high-redshift galaxies seeded with a MBH is about a factor of 10 below the direct collapse case presented in \cite{VLN2008}, where a seed was assumed to form with a 100\% efficiency whenever a protogalaxy disk was Toomre unstable.

\subsection{Primordial Black Holes}
\label{ssec:1d}
Primordial black holes may be formed also in the early universe by many  processes \citep[and references therein]{Carr2003}. Broadly speaking,  if within some region of space density  fluctuations are large, so that the gravitational force overcomes the 
pressure, the whole region can collapse and form a primordial black hole.  In the early Universe, generically, primordial black holes are formed with masses that roughly equal the mass within the particle horizon at the redshift of their formation \citep{Zeldovich1967,Hawking1971}. The masses of primordial black holes formed in the above mentioned 
processes range roughly from the Planck Mass (black holes formed at the Planck epoch) to $M_\odot$ (black holes formed at the QCD phase transition) up to $10^5\, \msun$ \citep{Russo2005}.

Several physical or astrophysical constraints restrict the mass range where primordial black holes are allowed. Primordial black holes with an initial mass smaller than about $5 \times 10^{14}$~g are  expected to be already evaporated due to Hawking radiation. For masses $\sim 10^{15}$~g, there are  strong bounds from the observed intensity of the diffuse gamma ray background \citep{Page1976}, limiting their contribution to the matter density to less than one part in $10^8$.  For 
larger masses, constraints can be deduced from microlensing  techniques \citep{Alcock2000,Tisserand2007} and from spectral distortions of the cosmic microwave background \citep{Ricotti2008} which limit the mass to below $\sim 10^3 \, \msun$. 

\section{The early growth of massive black holes}
Accretion is inevitable during the ``active" phase of a galactic nucleus. Observations tell us that AGN are widespread in both the local and early Universe. All  the information that we have gathered on the evolution of MBHs is indeed due to studies of AGN, as we have to await for the {\it Laser Interferometer Space Antenna} ({\it LISA}) to be able to ``observe" quiescent MBHs in the distant Universe. 

The classic argument of \cite{Soltan1982}, compares the total mass of black holes today with the total radiative output by known quasars,  by integration over redshift and luminosity of the luminosity function of quasars 
\citep{YuTremaine2002,Elvis2002,Marconi2004}.  The total energy density can  be converted  into the total mass density accreted by black holes during the active phase, by assuming a mass-to-energy conversion efficiency,  $\epsilon$ 
\citep{Aller2002,Merlonietal2004,Elvis2002,Marconi2004}:
\begin{equation}
\rho_{QSO}(<z)=\frac {f_{\rm bol}(1-\epsilon)} {\epsilon c^2}
\int_0^{z}\int{{{L' \Phi(L',z)} \over
{H_0(1+z)\sqrt{\Omega_m(1+z)^3 + \Omega_{\Lambda}}}}}{\rm d}L' {\rm d}z
\end{equation}
where the mass accretion rate,
$\dot M_{\rm acc}=f_{\rm bol}L\epsilon^{-1}c^{-2}$, is converted 
into MBH mass growth,
$\dot M_{\rm BH}=f_{\rm bol}(1-\epsilon)L\epsilon^{-1}c^{-2}$, with $f_{\rm bol}$ the bolometric
correction, and $\epsilon$ the energy conversion coefficient. 

The similarity of the total mass in MBHs today and the total mass accreted by MBHs  implies that the last 2-3 e-folds of the mass is grown via radiatively efficient accretion, rather than accumulated through mergers or radiatively inefficient accretion.  However,  most ot the Ôe-foldsÕ  (corresponding to a relatively small amount of mass, say the first  10\% of mass) could be gained rapidly, such as by, e.g.,  radiatively inefficient accretion.  This argument is particularly important at early times, since at least some MBHs must have grown rapidly. 

Luminous quasars have indeed been detected at very high redshift, $z>6$, when the Universe was less than {\rm 1 Gyr} old. Follow-up observations confirmed that at least some of  these quasars are powered by supermassive black holes with masses $\simeq 10^9\, M_\odot$ \citep{Barthetal2003,Willottetal2005}.   How could the seed massive black holes have grown rapidly enough within this short timespan is still an open question \citep[e.g.,][]{Haiman2004,Shapiro2005,VRees2006,Tanaka2009}.

The rate at which a seed massive black hole can accrete is limited by both `external' and `internal' effects. On the one hand, `external' conditions determine how much gas is available, at least initially, to the MBH.  Cosmological hydrodynamic simulations suggest that gravitational collapse produces dense central gas concentrations in protogalaxies \citep[e.g.,][]{pelupessy2007,Wise2008,Greif2008}. Atomic densities have been found to reach $n\sim 10^4\textrm{ cm}^{-3}$ \citep[e.g.,][]{Greif2008,BrommLoeb2003,Wise2008}.  On spatial scales that are resolved in the simulations, gas is sufficiently concentrated to enable rapid accretion onto a seed black hole, after the radiative feedback from the MBH progenitor star has ceased (if MBHs are the remnant of Population III stars, see \citealt{JBromm}).  

Given an ample gas supply, rapid accretion might be inhibited by `internal' feedback, that is the radiative output produced by the newly born quasar itself. \cite{pelupessy2007,JBromm} suggest that the growth of the MBH is severely limited by thermal feedback. However, existing cosmological simulations modeling MBH growth do not resolve the spatial scales that are needed to treat explicitly the radiative processes may alter accretion. The simulations also do not resolve the fine structure of the dense gaseous clouds where the MBHs are embedded. Sub-grid physics prescriptions are normally adopted for the accretion rate and the radiative feedback \citep[e.g.,][]{pelupessy2007,JBromm}. Recently, however, \cite{Milos2009} have simulated radiatively efficient accretion from a uniform, high-density metal-free protogalactic cloud onto a low-mass seed black hole. They found that a compact HII region forms around the black hole causing accretion to be intermittent, due to alternating radiation pressure-driven expulsion and external pressure-driven fallback. The average accretion rate is 1\% of the Bondi accretion rate calculated ignoring the radiation's influence, and 30\% of the Eddington-limited rate. 

This numerical experiment brings up the question on how black holes can handle sprees, that is if/when the supply is super-critical \citep{VolonteriRees2005} the excess radiation is effectively trapped.  At very high accretion rates\footnote{We distinguish here between super-Eddington, or super-critical, accretion rate and super-Eddington luminosity. A black hole accreting at super-critical values does not necessarily radiate at super-Eddington luminosity, depending on the properties of the accretion disc \citep{Abramowicz1988}.} radiation pressure cannot prevent the accretion rate from being super-critical, while the emergent luminosity is still Eddington limited in case of spherical or quasi-spherical configurations \citep{Begelman1979,BegelmanMeier1982}.  In the spherical case,  though this issue remains unclear, it still seems  possible that when the inflow rate is  super-critical, the radiative efficiency drops so that the hole can accept the material without  greatly exceeding the Eddington luminosity.  The  efficiency could be low either because most  radiation is trapped and advected inward, or because the flow adjusts so that the material can  plunge in from an orbit with small binding energy \citep{AbramowiczLasota1980}.  The creation of a radiation-driven outflow, which can possibly stop the infall of material, is also a possibility. 

The evidence for an accelerated growth of billion solar masses MBH in $z \simgt 6$ galaxies is stronger and stronger, though. 
\cite{Ajello2009} recently published the list  of blazars detected in the all sky survey by the  Burst Alert Telescope (BAT) onboard the {\it Swift} satellite. The derived luminosity function implies a space density $\sim 0.15$ Gpc$^{-3}$ in the redshift bin $z=3-4$ of blazars  pointing at us with hard X--ray  luminosities exceeding $2\times 10^{47}$ erg s$^{-1}$.  Using the IR-optical shape, that is produced by the accretion disk, whose radiation is
not beamed \citep{Ghisellini2009} find that in all cases of X-ray (beamed) luminosity exceeding $2\times 10^{47}$ erg s$^{-1}$, there is an IR-optical
(unbeamed) spectrum indicating MBHs with masses exceeding $10^9 M_\odot$ . Since blazars are selected  on the basis of their X--ray flux relativistically beamed at us, each of these objects implies the existence of other $\sim 2\Gamma^2$ misaligned similar sources. The space density of MBHs having $M>10^9 M_\odot$ in the $z=3-4$ redshift range is $\sim 68\, (\Gamma/15)^2$ Gpc$^{-3}$. The strong cosmological evolution revealed by the BAT blazars implies that beamed, jetted sources become dominant at the highest MBH masses ($>10^9 M_\odot$) and redshifts ($z>5$), when compared to the cosmic evolution of radio--quiet sources. 
The number density of blazars estimated from the BAT luminosity function is indeed comparable to the  \textsl{total} number density of $>10^9 M_\odot$ MBHs at $z>6$ possibly hosted in massive halos.  The presence of these astonishing sources suggests that normal `feedback' may not be at play at the highest redshifts. I expect that high-accretion rate events (distinctively possible during the violent early cosmic times) will trigger the formation of collimated outflows (e.g. blazars) that do not cause feedback directly on the host (which is pierced through), but will deposit their kinetic energy at large distances, leaving the host unscathed \citep[in a different context, see][]{Vernaleo2006}. This is likely if at large accretion rates photon trapping decreases the disk luminosity, while concurrently the presence of a jet helps dissipating angular momentum, thus promoting efficient accretion.  This suggestion may explain why high-redshift MBHs can accrete at very high rates without triggering self-regulation mechanisms.  

\section{Observational signatures of massive black hole seeds}

\begin{figure}
\includegraphics[width=0.75\textwidth]{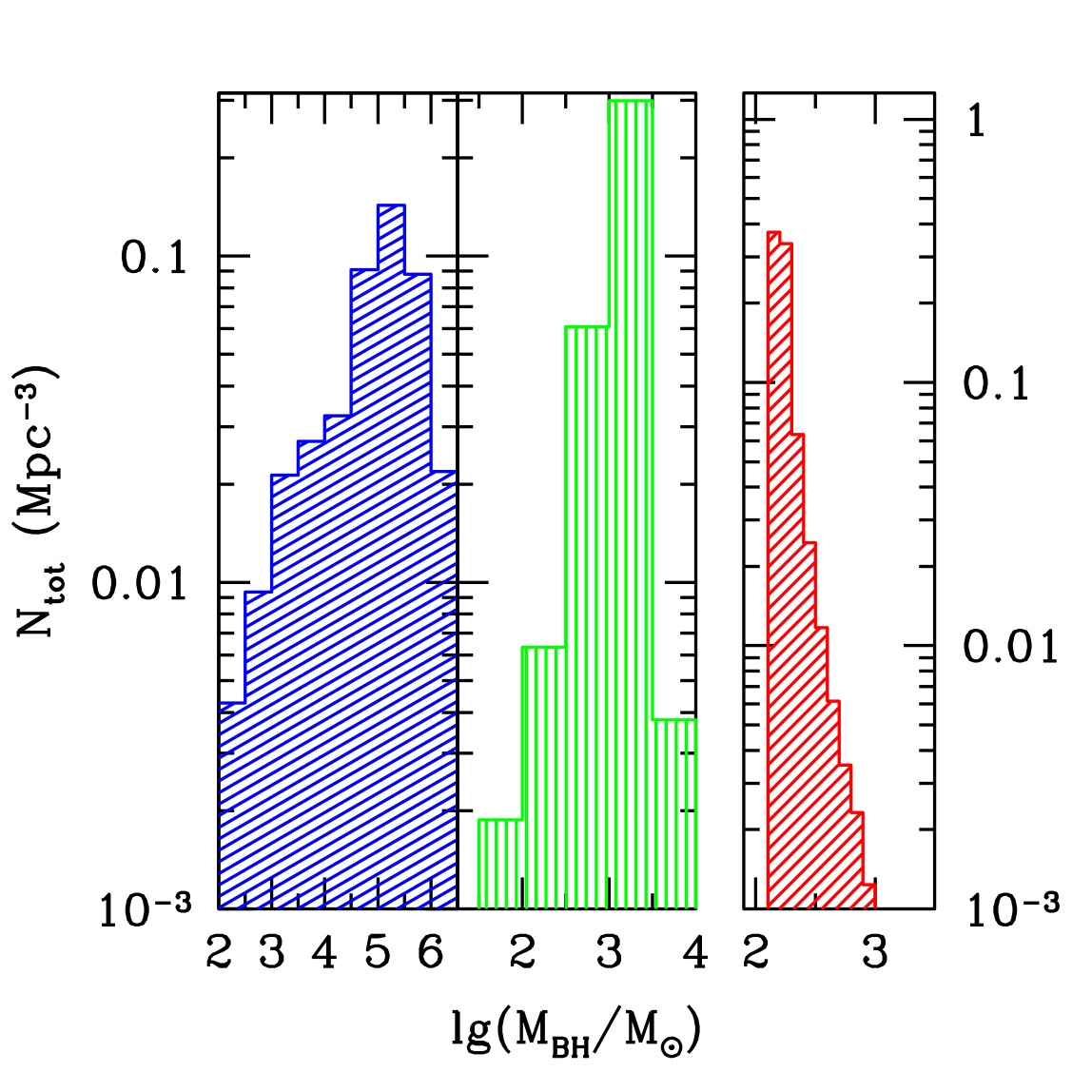}
\caption{Mass function of seed MBHs for three different formation scenarios: direct collapse \cite[left]{VLN2008}, runaway stellar mergers in high-redshift clusters \cite[center]{Devecchi2009}, and Population III remnants \cite[right]{MadauRees2001}. Note the different y-axis scale for the Population III case.}
\label{fig:MF}
\end{figure}

Figure~\ref{fig:MF} shows three mass functions for three different MBH `seed' models: direct collapse \citep{LN2006}, runaway stellar mergers in high-redshift clusters \citep{Devecchi2009}, and Population III remnants \citep{MadauRees2001}. These are the initial conditions that we would like to probe. In this section I will focus on two different, extreme, scenarios: `light seeds', derived from Population III remnants, and `heavy seeds', derived from direct gas collapse. In the following I delineate some possible observational tests that can be performed with current and future instruments. MBHs have unique properties that make them perfect ``beacons" at early cosmic times. MBHs that are active, that is, shining as quasars, are the most luminous sources in the Universe, making them lighthouses in the early stages of galaxy assembly, when the starlight from galaxies is too dim to be detected by our telescopes.  Also, black holes are not only seen in the electromagnetic spectrum, but they are also sirens for gravitational wave detectors, bringing about a novel way of observing our Universe.

\subsection{Tracing MBHs at the earliest times}
\begin{figure}   
   \includegraphics[width=\columnwidth]{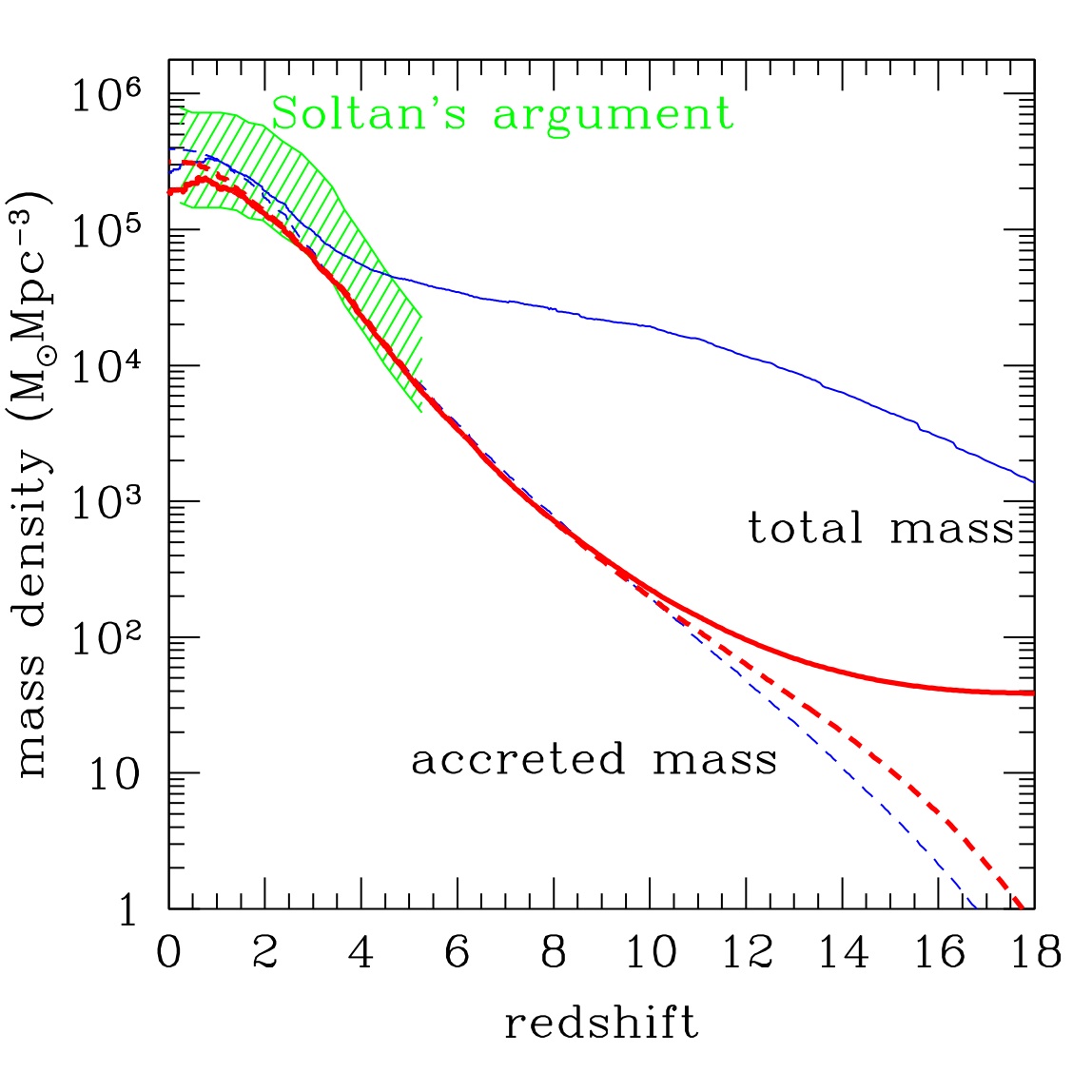} 
   \caption{Integrated black hole mass density as a function of
     redshift. Shaded area: constraints from Soltan-type  arguments, where we have varied the radiative efficiency from a
     lower limit of 6\% (upper    envelope of the shaded area), to about 30\%. Solid lines: total mass density locked into nuclear
     MBHs.  Dashed lines: integrated mass density {\rm  accreted} by MBHs.  Models based on remnants of
     Population III stars (lowest curve), models based on direct collapse \citep{LN2006}.  Notice that Soltan-type  arguments constrain the accreted mass density, rather than the total mass density.}
   \label{fig:rho}
\end{figure}

As MBHs increase their mass by a large factor  during the quasar epoch ($z\approx 3-4$)  \citep{YuTremaine2002}, signatures of the seed formation mechanisms are likely more evident at {\it earlier epochs}. Figure~\ref{fig:rho} compares the integrated comoving mass density in MBHs to the expectations from Soltan-type arguments, assuming that quasars are powered by radiatively efficient flows (for details, see \cite{Marconi2004,Hopkins2007}). The curves differ only with respect to the MBH formation scenario. We either assume that seeds are Population III remnants (red curve), or that seeds are formed via direct collapse \citep[blue]{LN2006}. We study MBH evolution within dark matter halos via a Monte-Carlo algorithm based on the extended Press-Schechter formalism. The population of MBHs evolves along with their hosts according to a ``merger driven scenario'', as described in \citep{VHM, VN09}. An accretion episode is assumed to occur  as a consequence of every major merger (mass ratio larger than 1:10) event.  During an accretion episode, each MBH accretes an amount of mass,  $\Delta M$, that scales with the $M_{\rm BH}-\sigma_*$ relation of its hosts \citep[see][]{VN09}. Accretion starts after a dynamical timescale and lasts until the MBH has accreted $\Delta M$.

While during and after the quasar epoch the mass densities in the theoretical models differ by less than a factor of 2 (`The night in which all cows are black', quoting the German philosopher Hegel), at $z>3$ the differences become more pronounced.  Notice that Soltan-type arguments constrain the accreted mass density (dashed curves in Figure~\ref{fig:rho}), rather than the total mass density (solid curves in Figure~\ref{fig:rho}), making the distinction much harder from the observational point of view.  

The initial conditions tend indeed to be erased very fast if accretion is efficient in growing MBHs. \cite{VG2009} focus on the ultra-high-redshift evolution of two distinct  populations of seed MBHs: `light seeds', derived from Population III remnants, and `heavy seeds', derived from direct gas collapse.  In both cases they treat accretion in a merger-driven fashion (as described above).  As a baseline model, one can assume that accretion proceeds at the Eddington rate \citep[see also][]{salvaterra07}. In a second case,  accretion during the active phase is based on the extrapolation of the  empirical distribution of Eddington ratios, $\lambda=\log(L_{\rm bol}/L_{\rm Edd})$, found in \cite{Merloni08}. We adopt a fitting function of the Eddington ratio distribution as a function of MBH mass and redshift \cite{Merloni08}.  We are here extrapolating such a model at much higher redshifts and lower MBH masses than originally intended. We caution readers in taking the results of this model face value. The main goal of this exercise is to probe possible sensible ranges for the accretion rates on MBHs. 

\begin{figure}[t]
\includegraphics[width=\textwidth]{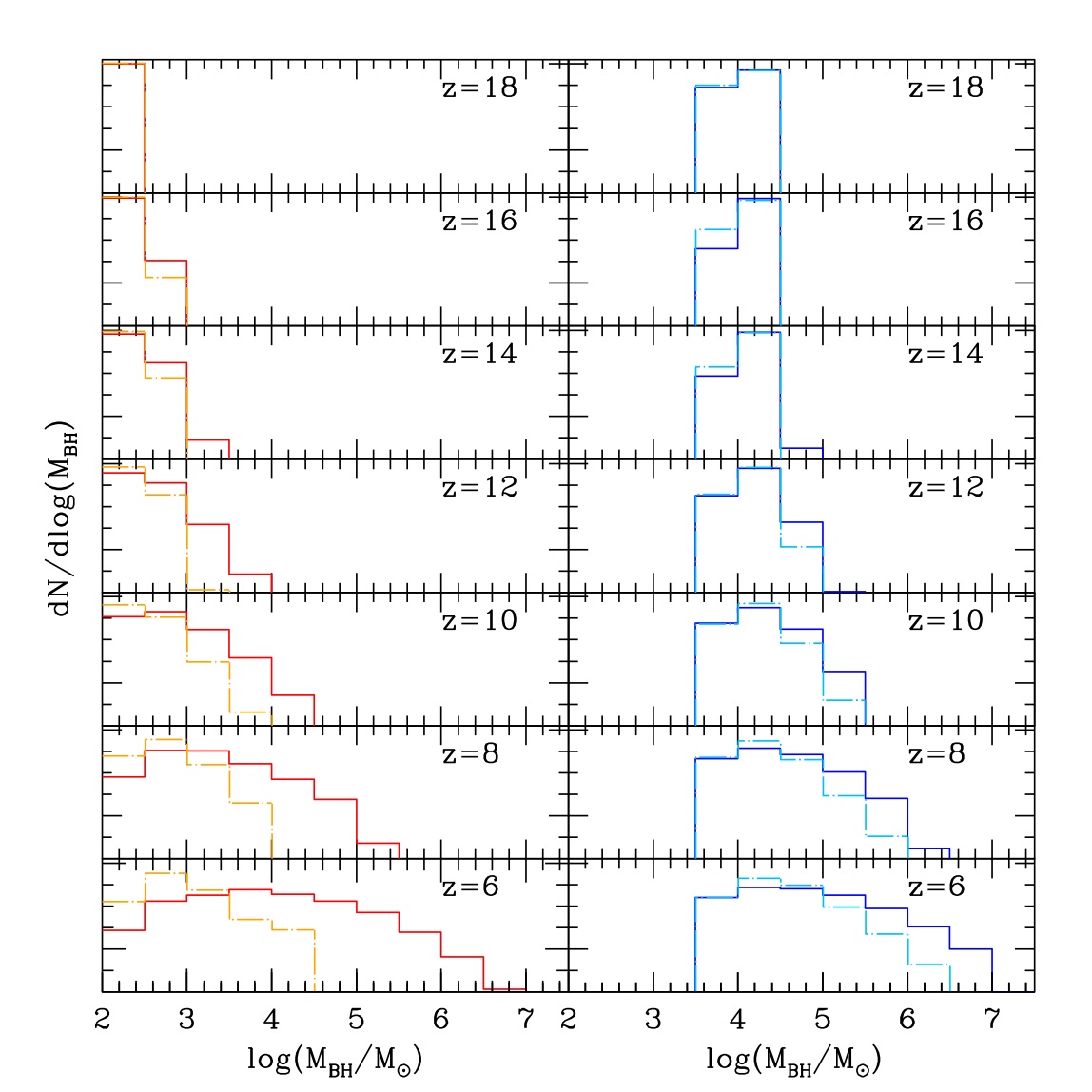}
\caption{Mass function of MBHs at different cosmic times. Top to bottom: $18<z<20$; $16<z<18$; $14<z<16$; $12<z<14$; $10<z<12$; $8<z<10$; $6<z<8$. Left panels: small seeds. Right panels: large seeds. Solid histogram: Eddington accretion; dot-dashed histogram: Merloni-Heinz accretion.  From \cite{VG2009}.}
\label{fig:bhmf}
\end{figure}

Figure~\ref{fig:bhmf} illustrates how, even starting from very different initial mass functions for the MBHs (top two panels) by $z=6$ the mass function at the largest masses is similar, if accretion is efficient. This washes out any trace of the initial seed population.  \cite{VG2009} also  find that of the contribution to the quasar luminosity budget  is dominated by MBHs  with mass $<10^6\msun$. Such small, low-luminosity MBHs do not contribute to the bright end of the luminosity function of  quasars, and are therefore difficult to account from simple extrapolations of the luminosity  function of quasars. These small holes are not hosted in extremely massive galaxies residing  in the highest density peaks (5 to 6--$\sigma$ peaks), but are instead found  in more common,  ``normal'' systems, $\sim3--\sigma$, peaks.  Future generation of space--based telescopes, such as {\it JWST} and {\it IXO}, are likely to detect and constrain the evolution of the population of accreting massive black holes at early times ($z\simgt10$). 

\subsection{Gravitational waves}
\begin{figure*} 
\includegraphics[width=0.75\textwidth]{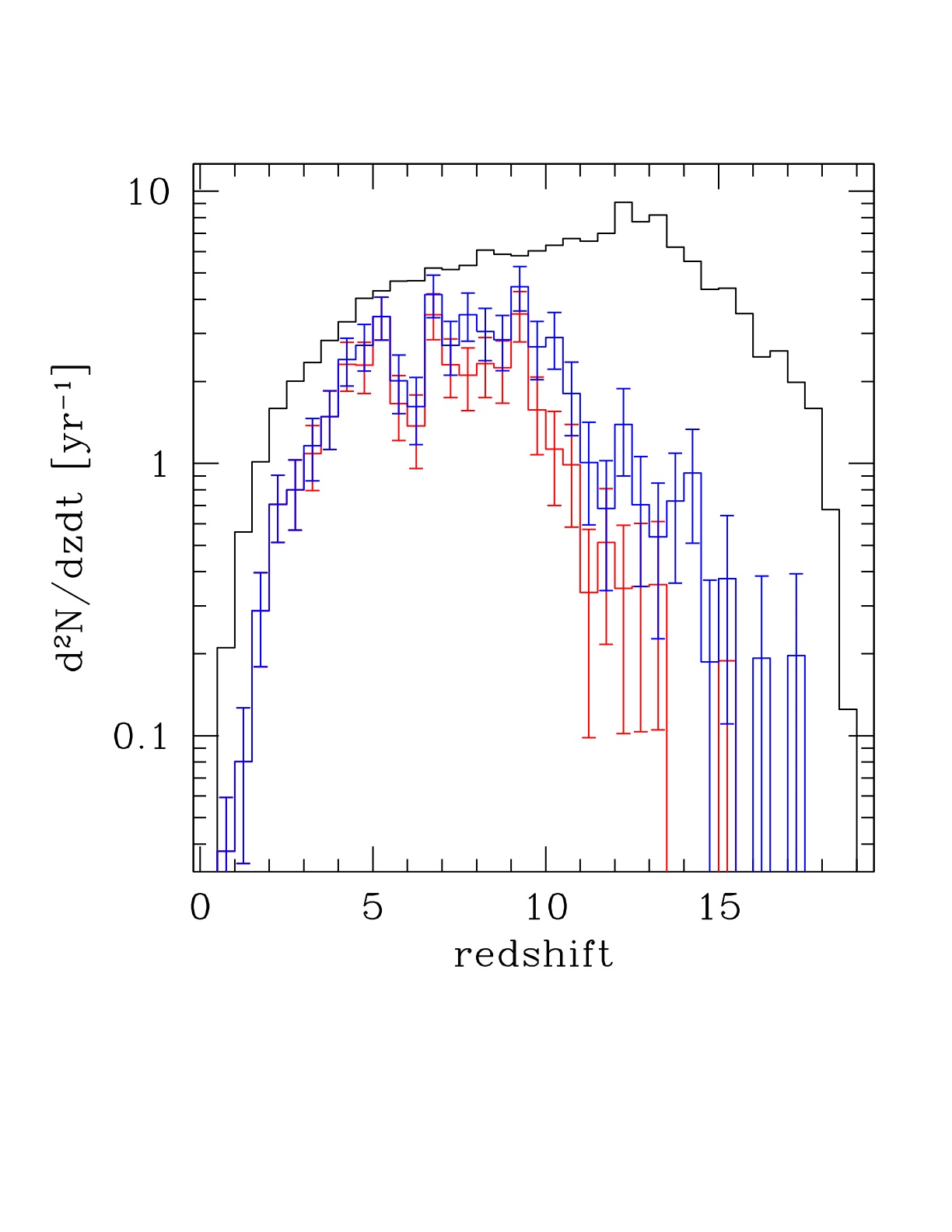}
\includegraphics[width=0.75\textwidth]{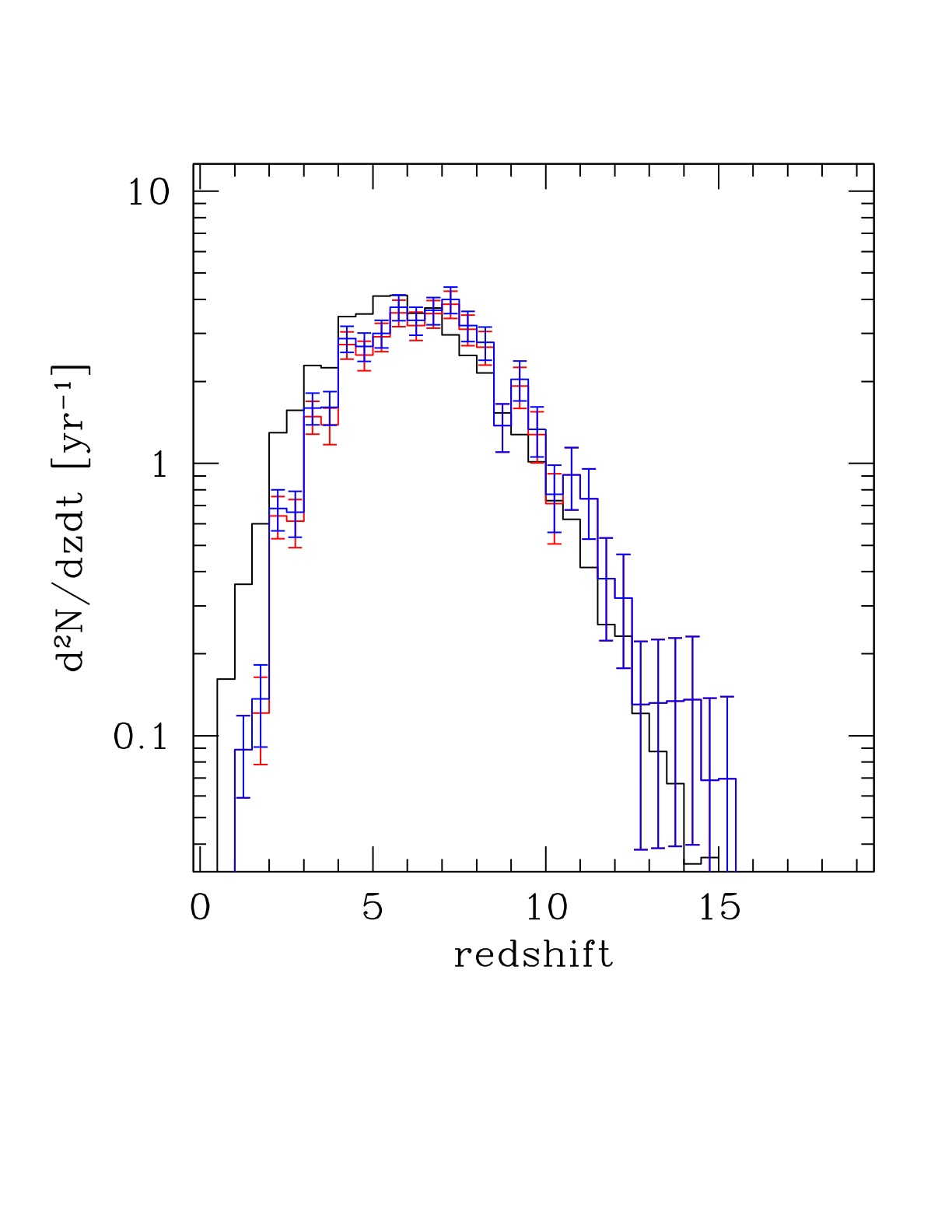}
  \caption{Merger rate of MBHs for two different seed models. Top: `light seeds' from population III remnants. Bottom: `heavy seeds' from gas-dynamical collapse. Black: all mergers.  Red: mergers detectable with S/N$>$10 in {\it LISA}'s baseline configuration. Blue:  mergers detectable with S/N$>$10 in the 6-link configuration. }
   \label{fig:lisa}
\end{figure*}

Detection of gravitational waves from seeds merging at the redshift of formation \citep{GW3} is probably one of the best ways to discriminate among formation mechanisms. {\it LISA} in principle is sensible to gravitational waves from binary MBHs with masses in the range $10^3-10^6\; M_\odot$ basically at any redshift of interest.   A large fraction of coalescences will be directly observable by {\it LISA}, and on the basis of the detection rate, constraints can be put on the MBH formation process. Different theoretical models for the formation of MBH seeds and dynamical evolution of the binaries indeed predict merger rates that largely vary one from the other (Figure~\ref{fig:lisa}).   

\cite{GW3} and \cite{arun2009} analyze merger histories of MBHs along the  hierarchical build--up of cosmic structures \citep{VHM}.  They consider two main scenarios for MBH formation \citep{MadauRees2001,BVR2006}.   \cite{GW3}  find that a decisive diagnostic is provided by the distribution  of the mass ratios in binary  coalescences. Models where seeds start large predict that most  of the detected events involve equal mass binaries. A fraction of observable coalescences, in fact, involve MBHs at $z>10$, when MBHs had no time to accrete  much mass yet. As most seeds form with similar mass, mergers at early times involve MBH binaries with mass ratio $\simeq 1$.  In scenarios based on Population III remnants, $z>10$ mergers involve MBHs with mass below the {\it LISA} threshold. The detectable events happen at later  times, when MBHs have already experienced a great deal of mass growth yielding a mass ratio distribution which is flat or features a broad  peak at mass ratios $\simeq 0.1-0.2$.   This is because both the probability of halo mergers (because of the steep host mass function) and the dynamical friction timescale increase with decreasing halo mass ratio. Hence, equal mass mergers that lead to efficient binary formation within short timescales (i.e., shorter than the Hubble time) are rare,  while in more common unequal mass mergers it takes longer than an Hubble time to drag  the satellite hole to the center. Finally, a  further, helpful diagnostic of black hole formation models lies in the shape of the mass distribution of detected events with S/N$>$5. Light binaries ($m<10^3\msun$) are predicted in a  fairly large number in Population  III  remnant models, but  are totally absent in direct collapse models.  

\cite{arun2009} determine the detectability of events by using  a code that includes both spin precession  and higher harmonics in the gravitational-wave signal, and carrying out Monte Carlo simulations to determine the number of events that can be detected  and accurately localized in these population models.   LISA will detect a significant fraction of the mergers which occur in the universe: almost all mergers will be detected in heavy-seed scenarios, and nearly half of all mergers in small-seed scenarios. This is because heavy seed black holes lead to more massive MBH binaries, so they can be seen out to larger redshifts.  Additionally, heavy-seed models are likely to produce more symmetric similar-mass binaries (which produce larger signal-to-noise ratios for fixed total mass). In contrast, light-seed models lead to more asymmetric systems.  Thus, although light-seed MBH mergers could occur more frequently, a smaller fraction would be observed by LISA due to their smaller total mass and less symmetric mass ratios.   For heavy-seed models most mergers detectable with high signal-noise-ratio (S/N$>$10) occur in the redshift range $3\simlt z \simlt 8$, with a peak around $z\sim 5$ (Figure\ \ref{fig:lisa}). In the case of light seeds, mergers are roughly uniform in $z$ over the range $4 \simlt z \simlt 10$.  Results are summarized in Table 1.

\begin{table}
  \caption{\cite{arun2009} evaluated parameter accuracies including also MBH spin information. The four models are referred to as follows. SE: Population III remnant seeds, high spins; SC: Population III remnant seeds, low spins; LE: gas-dynamical collapse, high spins; LC: gas-dynamical collapse, high spins.   For each merger-tree model (SE, SC, LE and LC) we list: the total
    number $N$ of merger events in LISA's past light cone in a one-year
    observation; the number of events $N_{\rm det}$ detectable with SNR larger
    than 10 in one year; the number for which the error in the luminosity
    distance $D_L$ is 10\% or less; the number that is localizable within 1
    and 10 deg$^2$ ($N_{\rm 1\, deg^2}$ and $N_{\rm 10\, deg^2}$,
    respectively);  and the    same for 1 deg$^2$ in angular resolution. Results are shown for two different LISA configurations and associated noise
curves: a ``6-link'' configuration, which allows the construction of all three
independent TDI channels, and a ``baseline'' configuration of 4 links,
producing a single Michelson channel. }
\label{tab:PE}
\begin{center}
\begin{tabular}{cccccc}
\hline\noalign{\smallskip}
Model  & $N$  & $N_{\rm det}$ & $N_{10\% D_L}$ & $N_{\rm 10\,deg^2}$  & $N_{\rm 1\, deg^2}$  \\ 
\hline
SE     & 80   &  33 (25)          & 21 (8.0)       & 8.2 (1.5)              & 2.2 (0.6)    \\
SC     & 75   &  34 (27)          & 17 (4.4)       & 6.1 (0.4)       & 1.3 (0.1)   \\
\hline
LE     & 24   &  23 (22)          & 21 (7.7)       & 10 (0.8)          & 2.2 (0.1)  \\
LC     & 22   &  21 (19)          & 14 (4.3)       &  6.5 (0.5)          & 1.8 (0.04)   \\
\noalign{\smallskip}\hline
\end{tabular}
\end{center}
\end{table}

\subsection{Footprints on the $\msigma$ relation}

\cite{VN09} explore the establishment and evolution of the empirical correlation between black hole mass ($M_{\rm BH}$) and velocity dispersion ($\sigma$) with redshift for two seeding models: `light seeds', derived from Population III remnants, and `heavy seeds', derived from direct gas collapse. Even though the seeds themselves do not satisfy the $M_{\rm BH} - \sigma$ relation initially, the relation can be established and maintained at all times if self-regulating accretion episodes are associated with major mergers. The massive end of the $M_{\rm BH} - \sigma$ relation is established early, and lower mass MBHs migrate onto it as hierarchical merging proceeds. How MBHs migrate toward the relation depends critically on the seeding prescription. Light seeds initially lie well below the $M_{\rm BH} - \sigma$ relation, and MBHs can grow via steady accretion episodes unhindered by self-regulation. In contrast, for the heavy seeding model, MBHs are initially over-massive compared to the empirical correlation, and the host haloes assemble prior to kick-starting the growth of the MBH.   The slope and scatter of the relation at the low-mass end, however, appear to be a consequence of the seeding mechanism and the self-regulation prescription. We find that if MBH seeds are massive, $\sim 10^5\,M_{\odot}$, the low-mass end of the $\msigma$ flattens towards an asymptotic value, creating a characteristic `plume'. This `plume' consists of ungrown seeds, that merely continue to track the peak of the seed mass function down to late times.  

\begin{figure*} 
\includegraphics[width=0.75\textwidth]{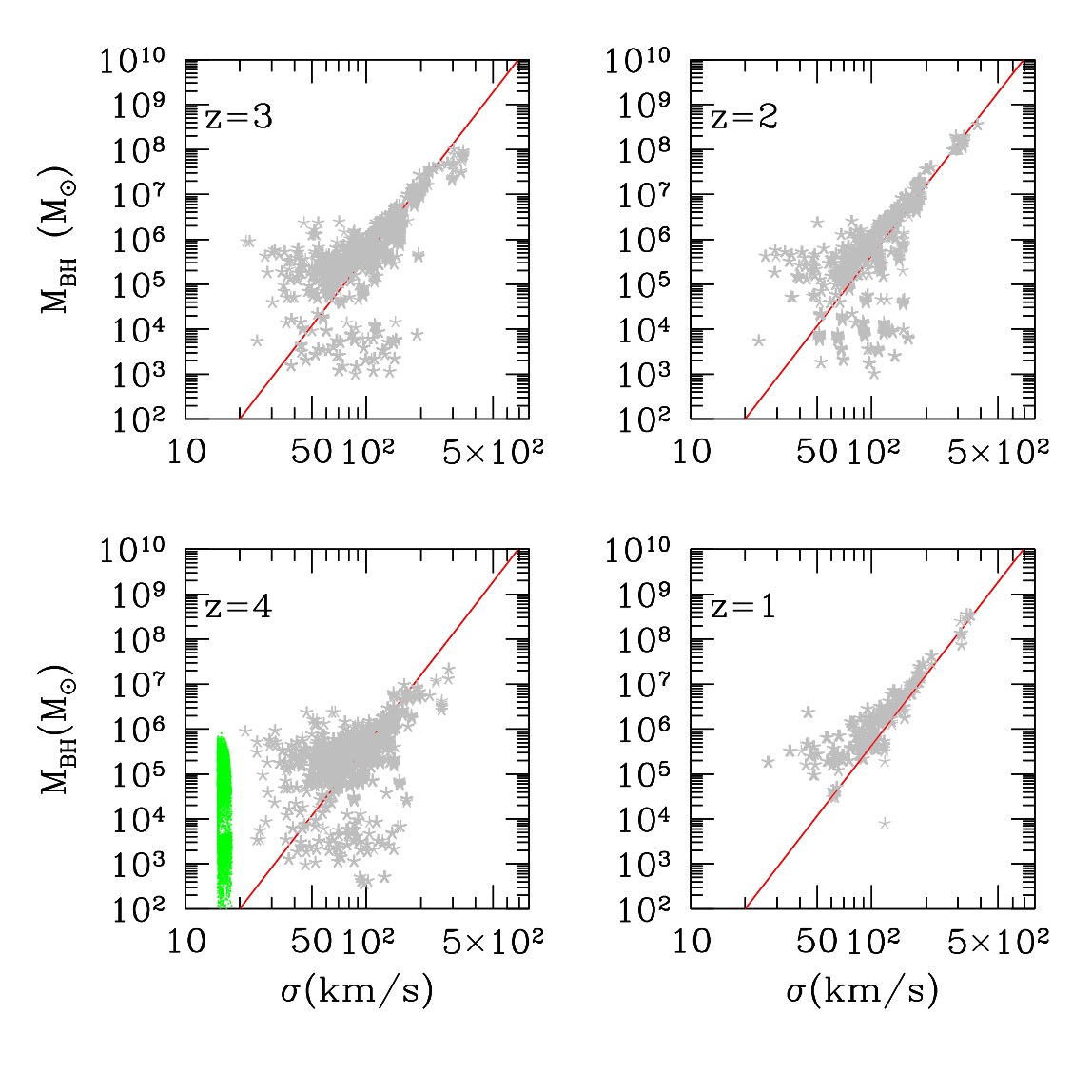}
\includegraphics[width=0.75\textwidth]{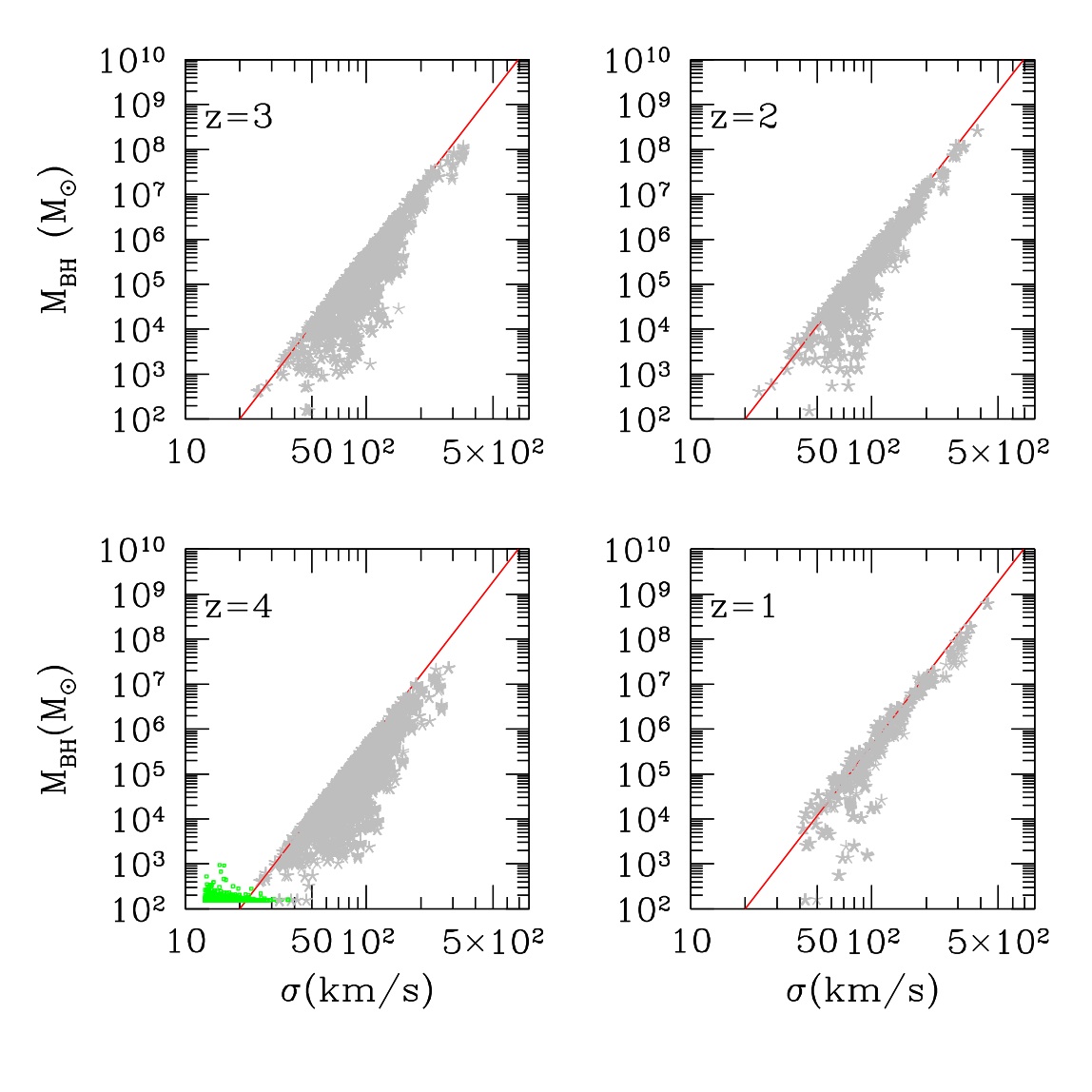}
  \caption{The $\msigma$ relation for MBHs at different redshifts. Top: MBHs evolve from an initial population of seeds based
  on the model by Lodato \& Natarajan (2006), with $Q_c=2$ (the lack   of any initial $\msigma$ correlation for seeds is clearly seen seen
  in the far left corner of the $z=4$ panels, green points).  Note the `plume' of MBHs at $\sigma
  <\,50\,\rm{km\,s}^{-1}$ that clearly persists even at $z = 2$ from   the earliest epochs. Bottom: $\msigma$ relation for the case of Population III remnant
seeds. The lack of an initial $\msigma$ correlation for these seeds is also evident here and is shown at the bottom of the $z=4$ panels (green points). From \cite{VN09}.}
   \label{fig:msigmaz}
\end{figure*}

Since it is during accretion episodes  that MBHs move  onto the $M_{\rm BH} - \sigma$ relation, AGN are better tracers  of the correlation itself, and worse tracers of the original seeds.  Additionally, differences between seeding models appear only at the low--mass end. That's because MBHs move on to the $\msigma$ relation  starting from the most massive systems at any time, as a consequence of major mergers being more common at high redshift for the most massive, biased, galaxies. The implication of this result is that flux limited AGN surveys tend to be biased toward finding MBHs that preferentially fall and anchor the $M_{\rm BH} - \sigma$ relation - these are systems where the initial conditions, the seed mass, have been erased.  One of the key predictions is the existence of a large population of low mass `hidden' MBHs at high redshift which are undetectable by flux limited AGN surveys. This population of low mass black holes ($M_{\rm BH} <10^6\,M_{\odot}$) are outliers at all epochs on the $M_{\rm BH} - \sigma$ relation, and they retain the strongest memory of the initial seed mass function, due to their quiet accretion history.  More accurate measurements of MBH masses below $M_{\rm BH} \sim 10^6\,M_{\odot}$ will enable us to use the measured $z = 0$ relation to constrain
seeding models at high redshift since cosmic evolution does not appear to smear out this imprint of the initial conditions. The scatter in the observed $\msigma$ relation might also provide insights into the initial seeding mechanism (Figure~\ref{fig:msigmaz}).

\subsection{Massive black holes in low-mass and dwarf galaxies}

The repercussions of different initial efficiencies for seed formation for the overall evolution of the MBH population stretch from high-redshift to the local Universe. The formation of seeds in a $\Lambda$CDM scenario follows the cosmological bias. As a consequence, the progenitors of massive galaxies (or clusters of galaxies) have a higher probability of hosting MBH seeds (cfr. \citealt{MadauRees2001}). In the case of low-bias systems, such as isolated low-mass galaxies, very few of the high-$z$ progenitors have the deep potential wells needed for gas retention and cooling, a prerequisite for MBH formation. 

\begin{figure*}   
\includegraphics[width=\textwidth]{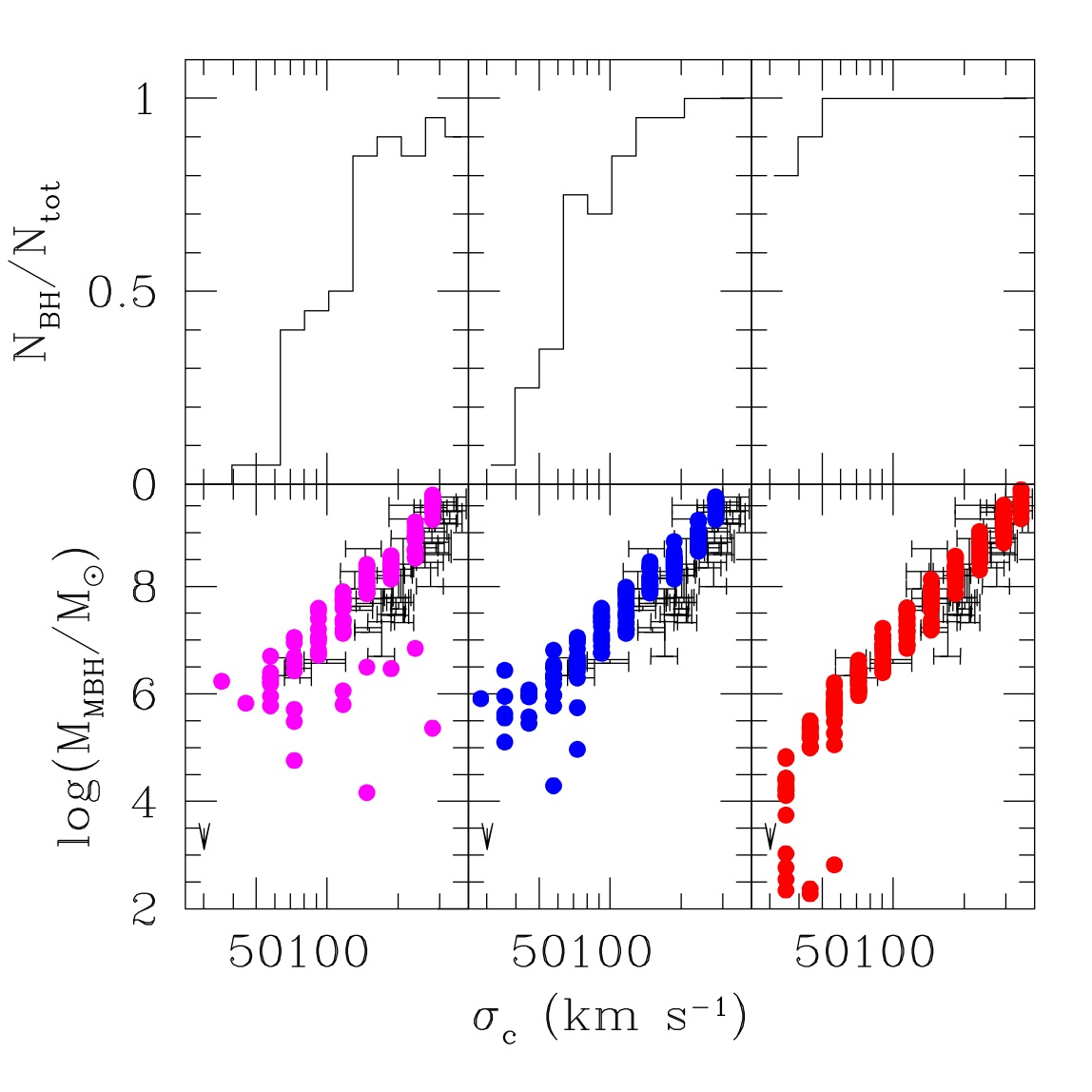}
   \caption{ The $M_{\rm bh}-$velocity dispersion ($\sigma_c$)
     relation at $z=0$. Every circle represents the central MBH in a
     halo of given $\sigma_c$.  Observational data are marked by their
     quoted errorbars, both in $\sigma_c$, and in $M_{\rm BH}$
     \citep{Tremaineetal2002}.  Left to right panels: $Q_{\rm c}=1.5$,
     $Q_{\rm c}=2$, Population III star seeds.  {\it
     Top panels:} fraction of galaxies at a given velocity dispersion
     which {\bf do} host a central MBH. From \cite{VLN2008}.}
   \label{fig3}
\end{figure*}

The signature of the efficiency of the formation of MBH seeds will consequently be stronger in low-mass galaxies. Figure~\ref{fig3}
(bottom panel) shows a comparison between the observed $M_{\rm BH}-\sigma$ relation and the one predicted by different models (shown with circles), and in particular, from left to right, two models based on the \citet{LN2006} seed masses with $Q_{\rm c}=1.5$ and 2 and a third model based on lower-mass Population III star seeds. The upper panel of Figure~\ref{fig3} shows the fraction of galaxies that  do host a massive black hole for different velocity dispersion bins. This shows that the fraction of galaxies with a MBH increases with increasing halo masses at $z = 0$.  A larger fraction of low mass halos are devoid of central black holes for lower seed formation efficiencies. This is one of the key discriminants between different formation scenarios.

These arguments can be applied to galaxies in different environments. Observationally, the best determination of the MBH (via AGN proxy) population has been performed in Virgo.  \citep{Decarli2007} analyzed nuclear activity in late type galaxies in the Virgo cluster using optical data. They conclude that at galaxy mass $M_{gal}> 10^{10.5}\msun$ the AGN fraction is unity. As a central black hole is a necessary condition for AGN activity, we conclude that the black hole occupation fraction must be unity as well. \cite{Gallo2008} conducted instead an X--ray survey where they find that the incidence of nuclear X-ray activity increases with the stellar mass  of the host galaxy: only between 3\% and 44\% of the galaxies with $M<10^{10} \msun$ harbor an X-ray active supermassive black hole. The fraction rises to between 49\% and 87\% in galaxies with stellar mass above $M<10^{10} \msun$.  \cite{Cote2006},  \cite{Wehner2006} and \cite{Ferrareseetal2006} find that, below $M_*\sim 10^{10}\msun$, Virgo spheroidals exhibit nuclear star clusters, whose mass scales with $M_*$  in the same fashion as those of the massive black holes detected in brighter galaxies \citep{Magorrian1998, MarconiHunt2003, Haring2004}.  The existence of a nuclear star cluster does not rule out the presence of a MBH, and nuclear star clusters coexist with MBHs over the whole range of MBH masses, such as in NGC 3115 \citep{Kormendy1996}. \cite{Ferrareseetal2006} however suggest that ``bright galaxies often, and perhaps always, contain supermassive black holes but not stellar nuclei. As one moves to fainter galaxies, nuclei become the dominant feature while  MBHs might become less common and perhaps disappear entirely at the faint end."   Our main interest here is to understand if MBHs populate preferentially bright galaxies, and if a mass threshold exists for a galaxy to host a MBH. 

The record for the smallest known MBH mass belongs to the dwarf Seyfert~1 galaxy POX 52. It is believed to contain a MBH of mass $M_{BH} \sim 10^5\,M_\odot$ \citep{barthetal2004}.  There are also significant non-detections of MBHs in a few nearby galaxies from stellar-dynamical observations, most notably the Local Group Scd-type spiral galaxy M33, in which the upper limit to $\mbh$ is just a few thousand solar masses \citep{Gebhardt2001,Merritt2001}.  Similarly, in the Local Group dwarf elliptical galaxy NGC 205, $\mbh < 3.8\times10^4 \msun$ \citep{Valluri2005}.  These results suggest that the MBH ``occupation fraction'' in low-mass galaxies might indeed be significantly below unity, but at present it is not possible to carry out measurements of similar sensitivity for galaxies much beyond the limits of the Local Group.

Indeed, simple arguments lead us to believe that MBHs might inhabit also the nuclei of dwarf galaxies, such as the satellites of the Milky Way and Andromeda, today. Indeed, one of the best diagnostics of `seed' formation mechanisms would be to measure the masses of MBHs in dwarf galaxies. As MBHs grow from lower-mass seeds, it is natural to expect that a leftover population of progenitor MBHs should also exist in the present universe.  As discussed above, the progenitors of massive galaxies have a high probability that the central MBH is not ``pristine", that is, it has increased its mass by accretion, or it has experienced mergers and dynamical interactions. Any dependence of $\mbh$ on the initial seed mass is largely erased.  Dwarf galaxies undergo a quieter merger history, and as a result, at low masses the MBH occupation fraction and the distribution of MBH masses still retain some ``memory'' of the original seed mass distribution. The signature of the efficiency of the formation of MBH seeds will consequently be stronger in  dwarf galaxies \citep{VLN2008}.

\begin{figure*}   
   \includegraphics[width=\columnwidth]{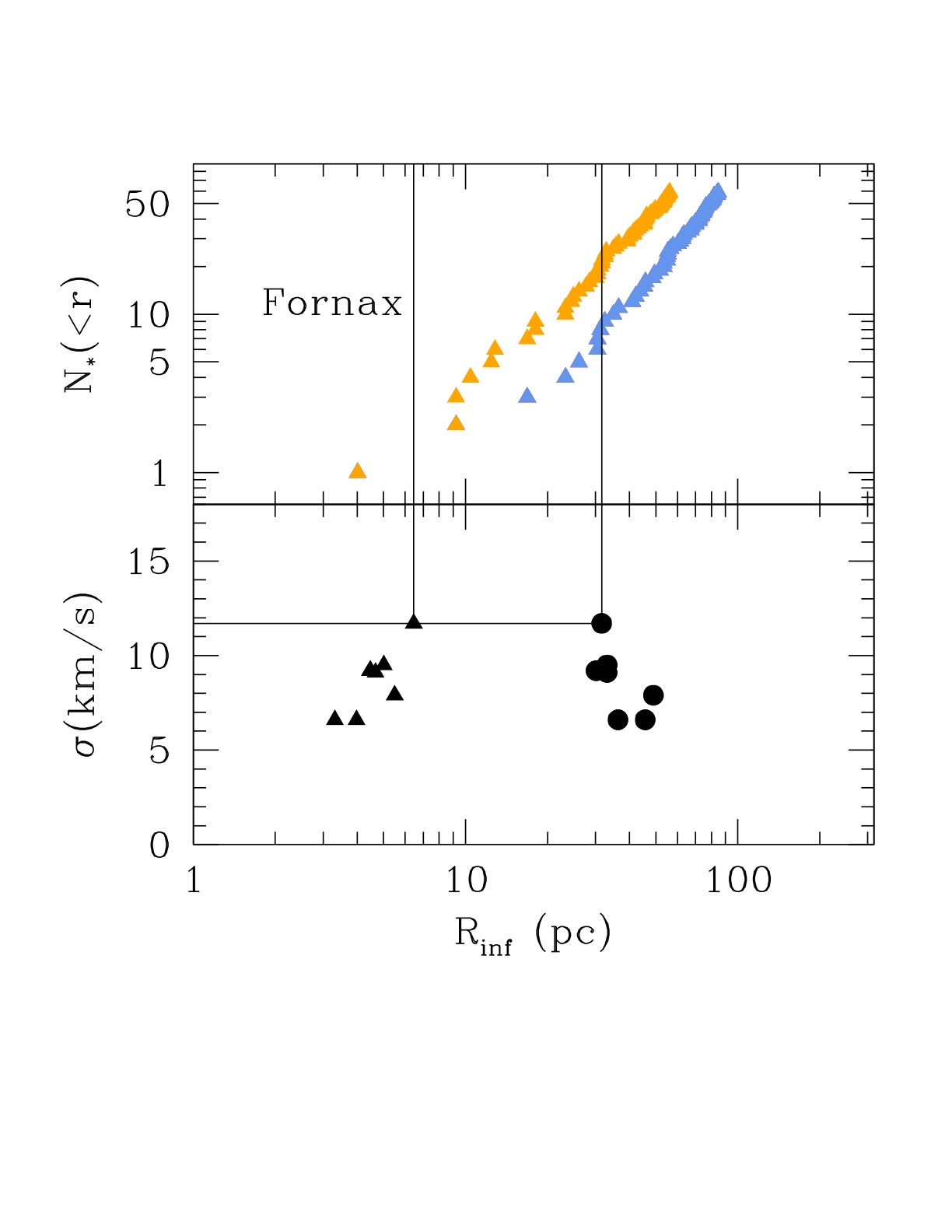} 
  \caption{Top panel: number of stars within a given projected radius in Fornax. Lower curve: all Fornax member stars for which velocities are currently available in the published kinematic samples of \cite{walker09a}. Upper curve:  all Fornax target candidates, including unobserved stars, that are sufficiently bright for velocity measurements with existing 6m -- 10m telescopes.   
      Bottom panel: relationship between velocity dispersion ($\sigma$) and radius of the sphere of influence of MBHs -- defined as the sphere that encompasses $2\times M_{BH}$ -- for ten halo realisations. 
      Triangles: we assume that the MBH sits on the $M_{BH}-\sigma$ relationship. Dots: we assume a fixed MBH mass, $10^5\msun$. Adapted from \cite{Svanwas2010}.}
   \label{fig:rinf}
\end{figure*}

\cite{Svanwas2010} find that for the most part MBHs hosted in Milky Way satellites retain the original `seed' mass, thus providing a clear indication of what the properties of the seeds were. MBHs generated as `heavy seeds' have larger masses, that would favour their identification, their typical occupation fraction is lower, being always below 40\% and decreasing to less than a \% for `true' dwarf galaxy sizes.  Light, Population III remnant, seeds have a higher occupation fraction, but their masses have not grown much since formation, making their detection harder.The presence of a quiescent MBH  can be tested dynamically if the region where the gravitational potential of the black hole dominates the gravitational potential of the host can be resolved. This region is referred to as the sphere of influence of the black hole. We adopt here the definition of the sphere of influence as the region within which the enclosed mass in stars equals twice the MBH mass. The radius of the sphere of influence is therefore defined as: $M(r<R_{\rm inf})=2\times M_{BH}$.

The lower panel in Figure~\ref{fig:rinf} plots stellar velocity dispersion against  the radius of the sphere of influence, $R_{\rm inf}$, estimated for  the eight ``classical'' dwarf spheroidal (dSph) satellites  of the Milky Way, for which line-of-sight velocities have been measured for up to a few thousand stars per galaxy \citep{walker09a}.  For these objects we adopt the stellar velocity dispersion measurements of \citet{Walker2009}, and then adopt an MBH mass from the mass-velocity dispersion relationship \citep{Tremaineetal2002}.  This case describes expectations likely for Pop III remnant seeds.  In order to calculate the sphere of influence for the dSphs, we consider the best-fitting mass profiles from \cite{Walker2009}.   In order to evaluate prospects for detecting kinematic signatures from such MBHs in real dSphs, the upper panel of Figure \ref{fig:rinf} indicates the number of spectroscopic target stars within a given projected radius in Fornax, the most luminous dSph satellite of the Milky Way.  Curves indicate the cumulative surface brightness profiles of 1) all Fornax member stars for which velocities are currently available in the published kinematic samples of Walker et al. 2009a , and 2) all Fornax target candidates, including unobserved stars, that are sufficiently bright  for velocity measurements with existing 6m -- 10m telescopes.  The latter profile represents the largest samples that are possible at present; unfortunately, these would include fewer than 5 stars within the spheres of influence estimated for the classical dSphs.  Thus even for the brightest nearby dSphs, the detection of any MBH must await the next generation of 20-30m telescopes, which may increase kinematic sample sizes by more than an order of magnitude.

Finally, we consider the spheres of influence due to MBH masses of $\sim 10^5\msun$, a mass of order of the upper limits derived for the `massive seed' scenario.  For such masses, the expected spheres of influence reach $\sim 50$ pc for the observed dsphs.  For Fornax, the expected value of $R_{\rm inf}\sim 30$ pc encloses 10 stars in the existing velocity sample, and 25 stars in the list of current target candidates.  If all these stars are observed, the resulting sample may help diagnose whether Fornax has an MBH of mass $\sim 10^5\msun$.

Pushing these limits further and probing the existence of MBHs in dwarf galaxies is observationally challenging.  An alternative way of detecting MBHs lurking in dwarf galaxies would be via their emission, when accreting surrounding material, either from a companion star or gas available as recycled material via mass loss of evolved stars (Dotti et al. in preparation).  A straightforward evidence for MBHs would in fact be the presence of AGNs \cite[and references therein]{Greene2007,Ho2008}. 
 One complication in the interpretation of AGN data is the possible contamination by X--ray binaries, that have a luminosity comparable to that expected from a $\sim 10^2$--$10^3\,M_\odot$ hole accreting from its surrounding gaseous environment.

Outside the local group, we can derive an estimate of expected number of dwarf galaxies which can possibly host MBHs through two different methods. First, we can rely on theoretical models of MBH formation and evolution, and look for the distribution of MBHs in dwarf galaxies. Using the dynamical model of \cite{Volonterietal08}, we estimate a number density of MBHs, $n_{\rm MBH}\sim 0.02$--$0.1$ Mpc$^{-3}$. 

Second, we can ground our estimate in recent theoretical works that study the population of dwarfs as satellites of the Milky Way \citep{Reed2005,Diemand2007,Springel2008}. These simulations suggest that the number of satellites per halo has the following form:
\beq
N(>v_{\rm sat})=N_*\left(\frac{v_{\rm sat}}{v_{\rm host}}\right) ^{\alpha},
\eeq
where $v_{\rm sat}$ and $v_{\rm host}$ are the maximum circular velocity of the satellite and the host halo, respectively. According to \cite{Diemand2007}, $N_*=0.021$ and $\alpha=-3$, while \cite{Springel2008} find $N_*=0.052$ and $\alpha=-3.15$.  If we extrapolate the $\msigma$ correlation to MBH masses ($10^2$--$10^3\,M_\odot$), and assume an isothermal galaxy, then $v_{\rm sat}\sim 10$--$20$ km~s$^{-1}$. With this formalism we obtain the number of satellites in the interesting mass range per dark matter halo ($N_{\rm sat}$), where the mass of the halo is uniquely determined by its maximum circular velocity.  Here, we use $N$ with appropriate subscripts to denote occupation number and $n$ to denote number density.  The number density of dark matter halos can be easily obtained by integrating the modified Press \& Schechter function \citep{Sheth1999} which provides the mass function of halos, $dn/dM_h$. Therefore we estimate a number density of satellites (per comoving cubic Mpc) as:
\beq
n_{\rm sat}=\int \frac{dn}{dM_h} N_{\rm sat}(M_h) dM_h=1-3 \, {\rm Mpc}^{-3},
\eeq
where the lower limit comes from \cite{Diemand2007}, and the upper limit from \cite{Springel2008}. 
We now have to correct for the fact that not all dwarf galaxies are likely to host an MBH, that is, the MBH ``occupation fraction'' is well below unity. To estimate the fraction of dwarfs that host a central MBH, we can rely on the models described in the previous paragraph, based on \citep{VHM, Volonterietal08}, in which a fraction $f_{\rm MBH}\sim 0.01$--$0.1$ of dwarfs host an MBH with mass $\sim 10^2$--$10^3\,M_\odot$. The final estimate for the number density of dwarfs hosting an MBH is then $n_{\rm MBH}=f_{\rm MBH} n_{\rm sat}\sim 0.01$--$0.3 \,{\rm Mpc}^{-3}$, in good agreement with the first estimate.

The detection of gravitational waves from a central MBH in a dwarf galaxy undergoing a merger is another possible probe.  The Einstein Telescope, a proposed third-generation ground-based gra\-vi\-ta\-tio\-nal-wave  detector  will be able to probe gravitational waves in a frequency range reaching down to $\sim 1$ Hz \citep{Freise:2009}. The frequency range determines the typical masses of coalescing binaries that could be detected by an interferometer; for example, the frequency of gravitational waves emitted from the innermost stable circular orbit of a test particle around a Schwarzschild black hole of mass $M$ is $\approx 4400\ {\rm Hz} (M_\odot/M)$.  The Einstein Telescope will therefore probe sources with masses of hundreds or a few thousand solar masses, which are out of reach of LISA or the current ground-based detectors.  Since dwarf galaxies  have a very quiet merger history, we do not expect many MBH-MBH mergers involving dwarf galaxies at the present epoch, or in the low--redshift universe. However, gravitational waves may also be generated in dwarf galaxies by mergers between the central MBH and stellar remnants (stellar mass BHs) in the centre of the dwarf. The core stellar densities in nearby dwarf galaxies are typically very low, e.g., the estimate for Fornax is $\sim 10^{-1}$pc$^{-3}$~\cite{Mateo:1998} and for Sagittarius is $\sim 10^{-3}$pc$^{-3}$~\cite{Majewski:2005}. When we calculate the event rate of BH-MBH mergers in dwarf galaxies, we have to further correct for the fact that only a small fraction of these tiny satellites do indeed form stars \cite[and references therein]{Bovill2009}. Based on \cite{Gnedin2006}, we estimate that a fraction $f_*=0.1-0.2$ of dwarfs in the  $v_{\rm sat}\sim 10-20$ km~s$^{-1}$ range formed stars (which will eventually leave behind stellar mass BHs that can merge with the central MBH).  The number density of MBHs that can be Einstein Telescope sources is therefore $n_{ET}=f_*\,n_{\rm MBH} \sim 0.001$--$0.06$ Mpc$^{-3}$. Therefore, although it is not inconceivable that the Einstein Telescope will detect events from dwarf galaxies, any events would be serendipitous \citep[$\ll 1$ per year,][]{Gair2009}. Detailed calculations are needed to understand/prove the robustness of these expectations.

\section{Conclusions}

We can trace the presence of `super' MBHs at early cosmic times, as the engines powering the luminous quasars that have been detected at high redshifts,  corresponding to  about 1 billion years after the Big Bang \citep{Fanetal2001a}.  The {\it HST}, {\it Chandra} and {\it Spitzer} satellites, jointly with 8-m class telescopes and large surveys, have made important breakthroughs, and observational cosmology probed capable of putting constraints on MBHs, when shining as quasars, up to high redshift \citep{Fanetal2001a, Fanetal2001b, Fanetal2004, Barthetal2003, Willottetal2003, Walteretal2004}.  Some constraints on the global accretion history, even at high redshift,  can be already put by comparing theoretical models predictions to ultra-deep X-ray surveys \citep{salvaterra07}. The early evolution of MBHs, and most notably, what physical mechanism is responsible for their formation are however still unknown. We now do know that MBHs {\it are} there, but we do not know how they {\it got} there. 

In this article I focused on three plausible mechanisms of MBH seed formation. Broadly speaking, we can divide them into physically-based categories:  `light seeds', forming at very early cosmic times ($M_{BH} \simeq 100-600 \msun$, $z\simeq 20-50$), `heavy seeds', forming later on ($M_{BH} \simeq 10^4-10^6 \msun$, $z\simeq 5-10$), and `intermediate seeds', forming with masses, and at epochs, in between the two previous cases ($M_{BH} \simeq 10^3 \msun$, $z\simeq 10-15$),. Light seeds forming early have a longer time to grow by accretion and MBH-MBH mergers, on the other hand, their accretion rates could be depressed in the shallow potential wells of the (mini-)halos, especially in the presence of radiative feedback from the (mini-)quasar itself \citep[and references therein]{Milos2009}. Heavy seeds forming later, in more substantial galaxies, are not likely to suffer from the same problems, but have had less time to grow. This would be partly compensated by their larger initial masses (Volonteri \& Begelman 2010).  At the current time observational constraints are too weak to favor one model against the others, but future X--ray missions,  such as {\it IXO}, and near infrared  facilities such as {\it JWST}, will have the technical capabilities to detect accreting MBHs at $z\gta 6$, giving constraints on the accretion properties of MBHs at early times. If the mass of the seeds is below $\sim10^5  \msun$, their flux is too weak for single sources to be detected electromagnetically.  Seeds of mass $\lta 10^5 \msun$ can nevertheless be directly identified during their mergers, by detecting their emission of gravitational radiation \citep{Hughes2002,berti2006}. Additionally, gravitational waves produced during the inspirals of compact objects into MBHs -- extreme-mass-ratio inspirals (EMRIs) are expected to provide accurate constraints on the population of MBHs in the $10^4M_{\odot}$--$10^7M_{\odot}$ range (Gair et al. 2010), which is the mass range where we can expect `memory' of the initial conditions, as detailed in section 4.4.  The combination of electromagnetic and gravitational wave observations in the coming years will improve the currently limited constraints on what route, or routes, lead to MBH seed formation. 

\begin{acknowledgements}

I wish to acknowledge here my long--time collaborators and mentors with whom I enjoyed many scientific discussions, Mitch Begelman, Francesco  Haardt, Piero Madau, Martin Rees. I am also grateful to all my more recent collaborators for putting up with my sometimes insane work rhythms. I promise I am a much more relaxed and pleasant person when I sit on a sandy beach with a book in my hands, the sun shining, and the ocean glistening.
\end{acknowledgements}


\begin{thebibliography}{165}
\providecommand{\natexlab}[1]{#1}
\providecommand{\url}[1]{{#1}}
\providecommand{\urlprefix}{URL }
\expandafter\ifx\csname urlstyle\endcsname\relax
  \providecommand{\doi}[1]{DOI~\discretionary{}{}{}#1}\else
  \providecommand{\doi}{DOI~\discretionary{}{}{}\begingroup
  \urlstyle{rm}\Url}\fi
\providecommand{\eprint}[2][]{\url{#2}}

\bibitem[{{Abel} et~al(2000){Abel}, {Bryan}, and {Norman}}]{abel2000}
{Abel} T, {Bryan} GL, {Norman} ML (2000) {The Formation and Fragmentation of
  Primordial Molecular Clouds}. {ApJ} 540:39--44, \doi{10.1086/309295},
  \eprint{astro-ph/0002135}

\bibitem[{{Abramowicz} and {Lasota}(1980)}]{AbramowiczLasota1980}
{Abramowicz} MA, {Lasota} JP (1980) {Spin-up of black holes by thick accretion
  disks}. Acta Astronomica 30:35--39

\bibitem[{{Abramowicz} et~al(1988){Abramowicz}, {Czerny}, {Lasota}, and
  {Szuszkiewicz}}]{Abramowicz1988}
{Abramowicz} MA, {Czerny} B, {Lasota} JP, {Szuszkiewicz} E (1988) {Slim
  accretion disks}. \apj 332:646--658, \doi{10.1086/166683}

\bibitem[{{Ajello} et~al(2009){Ajello}, {Costamante}, {Sambruna}, {Gehrels},
  {Chiang}, {Rau}, {Escala}, {Greiner}, {Tueller}, {Wall}, and
  {Mushotzky}}]{Ajello2009}
{Ajello} M, {Costamante} L, {Sambruna} RM, {Gehrels} N, {Chiang} J, {Rau} A,
  {Escala} A, {Greiner} J, {Tueller} J, {Wall} JV, {Mushotzky} RF (2009) {The
  Evolution of Swift/BAT Blazars and the Origin of the MeV Background}. \apj
  699:603--625, \doi{10.1088/0004-637X/699/1/603}, \eprint{0905.0472}

\bibitem[{{Alcock} et~al(2000){Alcock}, {Allsman}, {Alves}, {Axelrod},
  {Becker}, {Bennett}, {Cook}, {Dalal}, {Drake}, {Freeman}, {Geha}, {Griest},
  {Lehner}, {Marshall}, {Minniti}, {Nelson}, {Peterson}, {Popowski}, {Pratt},
  {Quinn}, {Stubbs}, {Sutherland}, {Tomaney}, {Vandehei}, and
  {Welch}}]{Alcock2000}
{Alcock} C, {Allsman} RA, {Alves} DR, {Axelrod} TS, {Becker} AC, {Bennett} DP,
  {Cook} KH, {Dalal} N, {Drake} AJ, {Freeman} KC, {Geha} M, {Griest} K,
  {Lehner} MJ, {Marshall} SL, {Minniti} D, {Nelson} CA, {Peterson} BA,
  {Popowski} P, {Pratt} MR, {Quinn} PJ, {Stubbs} CW, {Sutherland} W, {Tomaney}
  AB, {Vandehei} T, {Welch} D (2000) {The MACHO Project: Microlensing Results
  from 5.7 Years of Large Magellanic Cloud Observations}. \apj 542:281--307,
  \doi{10.1086/309512}, \eprint{arXiv:astro-ph/0001272}

\bibitem[{{Aller} and {Richstone}(2002)}]{Aller2002}
{Aller} MC, {Richstone} D (2002) {The Cosmic Density of Massive Black Holes
  from Galaxy Velocity Dispersions}. \aj 124:3035--3041, \doi{10.1086/344484}

\bibitem[{{Arun} et~al(2009){Arun}, {Babak}, {Berti}, {Cornish}, {Cutler},
  {Gair}, {Hughes}, {Iyer}, {Lang}, {Mandel}, {Porter}, {Sathyaprakash},
  {Sinha}, {Sintes}, {Trias}, {Van Den Broeck}, and {Volonteri}}]{arun2009}
{Arun} KG, {Babak} S, {Berti} E, {Cornish} N, {Cutler} C, {Gair} J, {Hughes}
  SA, {Iyer} BR, {Lang} RN, {Mandel} I, {Porter} EK, {Sathyaprakash} BS,
  {Sinha} S, {Sintes} AM, {Trias} M, {Van Den Broeck} C, {Volonteri} M (2009)
  {Massive black-hole binary inspirals: results from the LISA parameter
  estimation taskforce}. Classical and Quantum Gravity 26(9):094,027--+,
  \doi{10.1088/0264-9381/26/9/094027}, \eprint{0811.1011}

\bibitem[{{Barkat} et~al(1967){Barkat}, {Rakavy}, and {Sack}}]{Barkat1967}
{Barkat} Z, {Rakavy} G, {Sack} N (1967) {Dynamics of Supernova Explosion
  Resulting from Pair Formation}. Physical Review Letters 18:379--381,
  \doi{10.1103/PhysRevLett.18.379}

\bibitem[{{Barth} et~al(2003){Barth}, {Martini}, {Nelson}, and
  {Ho}}]{Barthetal2003}
{Barth} AJ, {Martini} P, {Nelson} CH, {Ho} LC (2003) {Iron Emission in the z =
  6.4 Quasar SDSS J114816.64+525150.3}. {ApJL} 594:L95--L98,
  \doi{10.1086/378735}

\bibitem[{{Barth} et~al(2004){Barth}, {Ho}, {Rutledge}, and
  {Sargent}}]{barthetal2004}
{Barth} AJ, {Ho} LC, {Rutledge} RE, {Sargent} WLW (2004) {POX 52: A Dwarf
  Seyfert 1 Galaxy with an Intermediate-Mass Black Hole}. {ApJ} 607:90--102,
  \doi{10.1086/383302}, \eprint{astro-ph/0402110}

\bibitem[{{Baumgarte} and {Shapiro}(1999)}]{Baumgarte1999}
{Baumgarte} TW, {Shapiro} SL (1999) {Evolution of Rotating Supermassive Stars
  to the Onset of Collapse}. \apj 526:941--952, \doi{10.1086/308006},
  \eprint{arXiv:astro-ph/9909237}

\bibitem[{{Begelman}(1979)}]{Begelman1979}
{Begelman} MC (1979) {Can a spherically accreting black hole radiate very near
  the Eddington limit}. {MNRAS} 187:237--251

\bibitem[{{Begelman}(2009)}]{Begelman2009b}
{Begelman} MC (2009) {Evolution of supermassive stars as a pathway to black
  hole formation}. ArXiv e-prints \eprint{0910.4398}

\bibitem[{{Begelman} and {Meier}(1982)}]{BegelmanMeier1982}
{Begelman} MC, {Meier} DL (1982) {Thick accretion disks - Self-similar,
  supercritical models}. {ApJ} 253:873--896, \doi{10.1086/159688}

\bibitem[{{Begelman} and {Rees}(1978)}]{Begelman1978}
{Begelman} MC, {Rees} MJ (1978) {The fate of dense stellar systems}. \mnras
  185:847--860

\bibitem[{{Begelman} and {Shlosman}(2009)}]{Begelman2009}
{Begelman} MC, {Shlosman} I (2009) {Angular Momentum Transfer and Lack of
  Fragmentation in Self-Gravitating Accretion Flows}. \apjl 702:L5--L8,
  \doi{10.1088/0004-637X/702/1/L5}, \eprint{0904.4247}

\bibitem[{{Begelman} et~al(2006){Begelman}, {Volonteri}, and {Rees}}]{BVR2006}
{Begelman} MC, {Volonteri} M, {Rees} MJ (2006) {Formation of supermassive black
  holes by direct collapse in pre-galactic haloes}. {MNRAS} 370:289--298,
  \doi{10.1111/j.1365-2966.2006.10467.x}, \eprint{astro-ph/0602363}

\bibitem[{{Begelman} et~al(2008){Begelman}, {Rossi}, and
  {Armitage}}]{Begelman2008}
{Begelman} MC, {Rossi} EM, {Armitage} PJ (2008) {Quasi-stars: accreting black
  holes inside massive envelopes}. \mnras 387:1649--1659,
  \doi{10.1111/j.1365-2966.2008.13344.x}, \eprint{0711.4078}

\bibitem[{{Berti} et~al(2006){Berti}, {Cardoso}, and {Will}}]{berti2006}
{Berti} E, {Cardoso} V, {Will} CM (2006) {Gravitational-wave spectroscopy of
  massive black holes with the space interferometer LISA}. Phys Rev D
  73(6):064,030--+, \doi{10.1103/PhysRevD.73.064030},
  \eprint{arXiv:gr-qc/0512160}

\bibitem[{{Blandford} and {Begelman}(1999)}]{Blandford1999}
{Blandford} RD, {Begelman} MC (1999) {On the fate of gas accreting at a low
  rate on to a black hole}. \mnras 303:L1--L5,
  \doi{10.1046/j.1365-8711.1999.02358.x}, \eprint{arXiv:astro-ph/9809083}

\bibitem[{{Blandford} and {Begelman}(2004)}]{Blandford2004}
{Blandford} RD, {Begelman} MC (2004) {Two-dimensional adiabatic flows on to a
  black hole - I. Fluid accretion}. \mnras 349:68--86,
  \doi{10.1111/j.1365-2966.2004.07425.x}, \eprint{arXiv:astro-ph/0306184}

\bibitem[{{Bond} et~al(1984){Bond}, {Arnett}, and {Carr}}]{Bond1984}
{Bond} JR, {Arnett} WD, {Carr} BJ (1984) {The evolution and fate of Very
  Massive Objects}. \apj 280:825--847, \doi{10.1086/162057}

\bibitem[{{Bovill} and {Ricotti}(2009)}]{Bovill2009}
{Bovill} MS, {Ricotti} M (2009) {Pre-Reionization Fossils, Ultra-Faint Dwarfs,
  and the Missing Galactic Satellite Problem}. \apj 693:1859--1870,
  \doi{10.1088/0004-637X/693/2/1859}, \eprint{0806.2340}

\bibitem[{{Bromm} and {Loeb}(2003)}]{BrommLoeb2003}
{Bromm} V, {Loeb} A (2003) {Formation of the First Supermassive Black Holes}.
  {ApJ} 596:34--46, \doi{10.1086/377529}, \eprint{astro-ph/0212400}

\bibitem[{{Bromm} et~al(1999){Bromm}, {Coppi}, and {Larson}}]{bromm1999}
{Bromm} V, {Coppi} PS, {Larson} RB (1999) {Forming the First Stars in the
  Universe: The Fragmentation of Primordial Gas}. {ApJL} 527:L5--L8,
  \doi{10.1086/312385}, \eprint{astro-ph/9910224}

\bibitem[{{Bromm} et~al(2002){Bromm}, {Coppi}, and {Larson}}]{bromm2002}
{Bromm} V, {Coppi} PS, {Larson} RB (2002) {The Formation of the First Stars. I.
  The Primordial Star-forming Cloud}. {ApJ} 564:23--51, \doi{10.1086/323947},
  \eprint{astro-ph/0102503}

\bibitem[{{Bullock} et~al(2001){Bullock}, {Dekel}, {Kolatt}, {Kravtsov},
  {Klypin}, {Porciani}, and {Primack}}]{Bullock2001}
{Bullock} JS, {Dekel} A, {Kolatt} TS, {Kravtsov} AV, {Klypin} AA, {Porciani} C,
  {Primack} JR (2001) {A Universal Angular Momentum Profile for Galactic
  Halos}. ApJ 555:240--257, \doi{10.1086/321477}, \eprint{astro-ph/0011001}

\bibitem[{{Carr}(2003)}]{Carr2003}
{Carr} BJ (2003) {Primordial Black Holes as a Probe of Cosmology and High
  Energy Physics}. In: {Giulini} D, {Kiefer} C, {Laemmerzahl} C (eds) Quantum
  Gravity: From Theory to Experimental Search, Lecture Notes in Physics, Berlin
  Springer Verlag, vol 631, pp 301--321

\bibitem[{{Carr} et~al(1984){Carr}, {Bond}, and {Arnett}}]{CBA84}
{Carr} BJ, {Bond} JR, {Arnett} WD (1984) {Cosmological consequences of
  Population III stars}. {ApJ} 277:445--469, \doi{10.1086/161713}

\bibitem[{{Clark} et~al(2008{\natexlab{a}}){Clark}, {Glover}, and
  {Klessen}}]{Glover2008}
{Clark} PC, {Glover} SCO, {Klessen} RS (2008{\natexlab{a}}) {The First Stellar
  Cluster}. \apj 672:757--764, \doi{10.1086/524187}

\bibitem[{{Clark} et~al(2008{\natexlab{b}}){Clark}, {Glover}, and
  {Klessen}}]{clark2008}
{Clark} PC, {Glover} SCO, {Klessen} RS (2008{\natexlab{b}}) {The First Stellar
  Cluster}. ApJ 672:757--764, \doi{10.1086/524187}, \eprint{arXiv:0706.0613}

\bibitem[{{C{\^o}t{\'e}} et~al(2006){C{\^o}t{\'e}}, {Piatek}, {Ferrarese},
  {Jord{\'a}n}, {Merritt}, {Peng}, {Ha{\c s}egan}, {Blakeslee}, {Mei}, {West},
  {Milosavljevi{\'c}}, and {Tonry}}]{Cote2006}
{C{\^o}t{\'e}} P, {Piatek} S, {Ferrarese} L, {Jord{\'a}n} A, {Merritt} D,
  {Peng} EW, {Ha{\c s}egan} M, {Blakeslee} JP, {Mei} S, {West} MJ,
  {Milosavljevi{\'c}} M, {Tonry} JL (2006) {The ACS Virgo Cluster Survey. VIII.
  The Nuclei of Early-Type Galaxies}. \apjs 165:57--94, \doi{10.1086/504042},
  \eprint{arXiv:astro-ph/0603252}

\bibitem[{{Decarli} et~al(2007){Decarli}, {Gavazzi}, {Arosio}, {Cortese},
  {Boselli}, {Bonfanti}, and {Colpi}}]{Decarli2007}
{Decarli} R, {Gavazzi} G, {Arosio} I, {Cortese} L, {Boselli} A, {Bonfanti} C,
  {Colpi} M (2007) {The census of nuclear activity of late-type galaxies in the
  Virgo cluster}. MNRAS 381:136--150, \doi{10.1111/j.1365-2966.2007.12208.x},
  \eprint{arXiv:0707.0999}

\bibitem[{{Devecchi} and {Volonteri}(2009)}]{Devecchi2009}
{Devecchi} B, {Volonteri} M (2009) {Formation of the First Nuclear Clusters and
  Massive Black Holes at High Redshift}. \apj 694:302--313,
  \doi{10.1088/0004-637X/694/1/302}, \eprint{0810.1057}

\bibitem[{{Diemand} et~al(2007){Diemand}, {Kuhlen}, and {Madau}}]{Diemand2007}
{Diemand} J, {Kuhlen} M, {Madau} P (2007) {Formation and Evolution of Galaxy
  Dark Matter Halos and Their Substructure}. \apj 667:859--877,
  \doi{10.1086/520573}, \eprint{arXiv:astro-ph/0703337}

\bibitem[{{Dijkstra} et~al(2008){Dijkstra}, {Haiman}, {Mesinger}, and
  {Wyithe}}]{Dijkstra2008}
{Dijkstra} M, {Haiman} Z, {Mesinger} A, {Wyithe} JSB (2008) {Fluctuations in
  the high-redshift Lyman-Werner background: close halo pairs as the origin of
  supermassive black holes}. \mnras 391:1961--1972,
  \doi{10.1111/j.1365-2966.2008.14031.x}, \eprint{0810.0014}

\bibitem[{{Ebisuzaki} et~al(2001){Ebisuzaki}, {Makino}, {Tsuru}, {Funato},
  {Portegies Zwart}, {Hut}, {McMillan}, {Matsushita}, {Matsumoto}, and
  {Kawabe}}]{ebisuzaki2001}
{Ebisuzaki} T, {Makino} J, {Tsuru} TG, {Funato} Y, {Portegies Zwart} S, {Hut}
  P, {McMillan} S, {Matsushita} S, {Matsumoto} H, {Kawabe} R (2001) {Missing
  Link Found? The ``Runaway'' Path to Supermassive Black Holes}. ApJL
  562:L19--L22, \doi{10.1086/338118}, \eprint{astro-ph/0106252}

\bibitem[{{Eisenstein} and {Loeb}(1995)}]{Eisenstein1995}
{Eisenstein} DJ, {Loeb} A (1995) {Origin of quasar progenitors from the
  collapse of low-spin cosmological perturbations}. \apj 443:11--17,
  \doi{10.1086/175498}, \eprint{arXiv:astro-ph/9401016}

\bibitem[{{Eke} et~al(1996){Eke}, {Cole}, and {Frenk}}]{Eke1996}
{Eke} VR, {Cole} S, {Frenk} CS (1996) {Cluster evolution as a diagnostic for
  Omega}. \mnras 282:263--280, \eprint{arXiv:astro-ph/9601088}

\bibitem[{{Elvis} et~al(2002){Elvis}, {Risaliti}, and {Zamorani}}]{Elvis2002}
{Elvis} M, {Risaliti} G, {Zamorani} G (2002) {Most Supermassive Black Holes
  Must Be Rapidly Rotating}. \apjl 565:L75--L77, \doi{10.1086/339197},
  \eprint{arXiv:astro-ph/0112413}

\bibitem[{{Fan} et~al(2001{\natexlab{a}})}]{Fanetal2001b}
{Fan} X, et~al (2001{\natexlab{a}}) {A Survey of $z>5.8$ Quasars in the Sloan
  Digital Sky Survey. I. Discovery of Three New Quasars and the Spatial Density
  of Luminous Quasars at z$\sim$6}. {AJ} 122:2833--2849, \doi{10.1086/324111}

\bibitem[{{Fan} et~al(2001{\natexlab{b}})}]{Fanetal2001a}
{Fan} X, et~al (2001{\natexlab{b}}) {High-Redshift Quasars Found in Sloan
  Digital Sky Survey Commissioning Data. IV. Luminosity Function from the Fall
  Equatorial Stripe Sample}. {AJ} 121:54--65, \doi{10.1086/318033}

\bibitem[{{Fan} et~al(2004)}]{Fanetal2004}
{Fan} X, et~al (2004) {A Survey of $z>5.7$ Quasars in the Sloan Digital Sky
  Survey. III. Discovery of Five Additional Quasars}. {AJ} 128:515--522,
  \doi{10.1086/422434}

\bibitem[{{Ferrarese} and {Ford}(2005)}]{ferrareseford}
{Ferrarese} L, {Ford} H (2005) {Supermassive Black Holes in Galactic Nuclei:
  Past, Present and Future Research}. Space Science Reviews 116:523--624,
  \doi{10.1007/s11214-005-3947-6}, \eprint{astro-ph/0411247}

\bibitem[{{Ferrarese} and {Merritt}(2000)}]{fm00}
{Ferrarese} L, {Merritt} D (2000) {A Fundamental Relation between Supermassive
  Black Holes and Their Host Galaxies}. {ApJ} 539:L9--L12, \doi{10.1086/312838}

\bibitem[{{Ferrarese} et~al(2006){Ferrarese}, {C{\^o}t{\'e}}, {Dalla
  Bont{\`a}}, {Peng}, {Merritt}, {Jord{\'a}n}, {Blakeslee}, {Ha{\c s}egan},
  {Mei}, {Piatek}, {Tonry}, and {West}}]{Ferrareseetal2006}
{Ferrarese} L, {C{\^o}t{\'e}} P, {Dalla Bont{\`a}} E, {Peng} EW, {Merritt} D,
  {Jord{\'a}n} A, {Blakeslee} JP, {Ha{\c s}egan} M, {Mei} S, {Piatek} S,
  {Tonry} JL, {West} MJ (2006) {A Fundamental Relation between Compact Stellar
  Nuclei, Supermassive Black Holes, and Their Host Galaxies}. {ApJL}
  644:L21--L24, \doi{10.1086/505388}, \eprint{astro-ph/0603840}

\bibitem[{{Freese} et~al(2008){Freese}, {Bodenheimer}, {Spolyar}, and
  {Gondolo}}]{Freese2008}
{Freese} K, {Bodenheimer} P, {Spolyar} D, {Gondolo} P (2008) {Stellar Structure
  of Dark Stars: A First Phase of Stellar Evolution Resulting from Dark Matter
  Annihilation}. \apjl 685:L101--L104, \doi{10.1086/592685}, \eprint{0806.0617}

\bibitem[{{Freise} et~al(2009){Freise}, {Chelkowski}, {Hild}, {Del Pozzo},
  {Perreca}, and {Vecchio}}]{Freise:2009}
{Freise} A, {Chelkowski} S, {Hild} S, {Del Pozzo} W, {Perreca} A, {Vecchio} A
  (2009) {Triple Michelson interferometer for a third-generation gravitational
  wave detector}. Classical and Quantum Gravity 26(8):085,012--+,
  \doi{10.1088/0264-9381/26/8/085012}, \eprint{0804.1036}

\bibitem[{{Freitag} et~al(2006{\natexlab{a}}){Freitag}, {G{\"u}rkan}, and
  {Rasio}}]{freitag2006b}
{Freitag} M, {G{\"u}rkan} MA, {Rasio} FA (2006{\natexlab{a}}) {Runaway
  collisions in young star clusters - II. Numerical results}. MNRAS
  368:141--161, \doi{10.1111/j.1365-2966.2006.10096.x},
  \eprint{astro-ph/0503130}

\bibitem[{{Freitag} et~al(2006{\natexlab{b}}){Freitag}, {Rasio}, and
  {Baumgardt}}]{freitag2006a}
{Freitag} M, {Rasio} FA, {Baumgardt} H (2006{\natexlab{b}}) {Runaway collisions
  in young star clusters - I. Methods and tests}. MNRAS 368:121--140,
  \doi{10.1111/j.1365-2966.2006.10095.x}, \eprint{astro-ph/0503129}

\bibitem[{{Fryer} et~al(2001){Fryer}, {Woosley}, and {Heger}}]{fryer2001}
{Fryer} CL, {Woosley} SE, {Heger} A (2001) {Pair-Instability Supernovae,
  Gravity Waves, and Gamma-Ray Transients}. {ApJ} 550:372--382,
  \doi{10.1086/319719}, \eprint{astro-ph/0007176}

\bibitem[{{Gaburov} et~al(2009){Gaburov}, {Lombardi}, and {Portegies
  Zwart}}]{Gaburov2009}
{Gaburov} E, {Lombardi} J, {Portegies Zwart} S (2009) {On the onset of runaway
  stellar collisions in dense star clusters - II. Hydrodynamics of three-body
  interactions}. ArXiv e-prints \eprint{0904.0997}

\bibitem[{{Gair} et~al(2009){Gair}, {Mandel}, {Miller}, and
  {Volonteri}}]{Gair2009}
{Gair} JR, {Mandel} I, {Miller} MC, {Volonteri} M (2009) {Exploring
  intermediate and massive black-hole binaries with the Einstein Telescope}.
  ArXiv e-prints \eprint{0907.5450}

\bibitem[{{Gallo} et~al(2008){Gallo}, {Treu}, {Jacob}, {Woo}, {Marshall}, and
  {Antonucci}}]{Gallo2008}
{Gallo} E, {Treu} T, {Jacob} J, {Woo} J, {Marshall} PJ, {Antonucci} R (2008)
  {AMUSE-Virgo. I. Supermassive Black Holes in Low-Mass Spheroids}. \apj
  680:154--168, \doi{10.1086/588012}, \eprint{0711.2073}

\bibitem[{{Gao} et~al(2007){Gao}, {Yoshida}, {Abel}, {Frenk}, {Jenkins}, and
  {Springel}}]{gao2006}
{Gao} L, {Yoshida} N, {Abel} T, {Frenk} CS, {Jenkins} A, {Springel} V (2007)
  {The first generation of stars in the {$\Lambda$} cold dark matter
  cosmology}. MNRAS 378:449--468, \doi{10.1111/j.1365-2966.2007.11814.x},
  \eprint{arXiv:astro-ph/0610174}

\bibitem[{{Gebhardt} et~al(2000){Gebhardt}, {Bender}, {Bower}, {Dressler},
  {Faber}, {Filippenko}, {Green}, {Grillmair}, {Ho}, {Kormendy}, {Lauer},
  {Magorrian}, {Pinkney}, {Richstone}, and {Tremaine}}]{Gebhardt2000}
{Gebhardt} K, {Bender} R, {Bower} G, {Dressler} A, {Faber} SM, {Filippenko} AV,
  {Green} R, {Grillmair} C, {Ho} LC, {Kormendy} J, {Lauer} TR, {Magorrian} J,
  {Pinkney} J, {Richstone} D, {Tremaine} S (2000) {A Relationship between
  Nuclear Black Hole Mass and Galaxy Velocity Dispersion}. {ApJ} 539:L13--L16,
  \doi{10.1086/312840}, \eprint{astro-ph/0006289}

\bibitem[{{Gebhardt} et~al(2001){Gebhardt}, {Lauer}, {Kormendy}, {Pinkney},
  {Bower}, {Green}, {Gull}, {Hutchings}, {Kaiser}, {Nelson}, {Richstone}, and
  {Weistrop}}]{Gebhardt2001}
{Gebhardt} K, {Lauer} TR, {Kormendy} J, {Pinkney} J, {Bower} GA, {Green} R,
  {Gull} T, {Hutchings} JB, {Kaiser} ME, {Nelson} CH, {Richstone} D, {Weistrop}
  D (2001) {M33: A Galaxy with No Supermassive Black Hole}. \aj 122:2469--2476,
  \doi{10.1086/323481}, \eprint{arXiv:astro-ph/0107135}

\bibitem[{{Ghez} et~al(2005){Ghez}, {Salim}, {Hornstein}, {Tanner}, {Lu},
  {Morris}, {Becklin}, and {Duch{\^e}ne}}]{Ghez2005}
{Ghez} AM, {Salim} S, {Hornstein} SD, {Tanner} A, {Lu} JR, {Morris} M,
  {Becklin} EE, {Duch{\^e}ne} G (2005) {Stellar Orbits around the Galactic
  Center Black Hole}. \apj 620:744--757, \doi{10.1086/427175},
  \eprint{arXiv:astro-ph/0306130}

\bibitem[{{Ghisellini} et~al(2009){Ghisellini}, {Foschini}, {Volonteri},
  {Ghirlanda}, {Haardt}, {Burlon}, and {Tavecchio}}]{Ghisellini2009}
{Ghisellini} G, {Foschini} L, {Volonteri} M, {Ghirlanda} G, {Haardt} F,
  {Burlon} D, {Tavecchio} F (2009) {The blazar S5 0014+813: a real or apparent
  monster?} \mnras 399:L24--L28, \doi{10.1111/j.1745-3933.2009.00716.x},
  \eprint{0906.0575}

\bibitem[{{Glover} et~al(2008){Glover}, {Clark}, {Greif}, {Johnson}, {Bromm},
  {Klessen}, and {Stacy}}]{Glover2008a}
{Glover} SCO, {Clark} PC, {Greif} TH, {Johnson} JL, {Bromm} V, {Klessen} RS,
  {Stacy} A (2008) {Open questions in the study of population III star
  formation}. In: {Hunt} LK, {Madden} S, {Schneider} R (eds) IAU Symposium, IAU
  Symposium, vol 255, pp 3--17, \doi{10.1017/S1743921308024526}

\bibitem[{{Gnedin} and {Kravtsov}(2006)}]{Gnedin2006}
{Gnedin} NY, {Kravtsov} AV (2006) {Fossils of Reionization in the Local Group}.
  \apj 645:1054--1061, \doi{10.1086/504404}, \eprint{arXiv:astro-ph/0601401}

\bibitem[{{Greene} and {Ho}(2007)}]{Greene2007}
{Greene} JE, {Ho} LC (2007) {The Mass Function of Active Black Holes in the
  Local Universe}. \apj 667:131--148, \doi{10.1086/520497}, \eprint{0705.0020}

\bibitem[{{Greif} et~al(2008){Greif}, {Johnson}, {Klessen}, and
  {Bromm}}]{Greif2008}
{Greif} TH, {Johnson} JL, {Klessen} RS, {Bromm} V (2008) {The first galaxies:
  assembly, cooling and the onset of turbulence}. \mnras 387:1021--1036,
  \doi{10.1111/j.1365-2966.2008.13326.x}, \eprint{0803.2237}

\bibitem[{{Gruzinov}(1998)}]{Gruzinov1998}
{Gruzinov} AV (1998) {Radiative Efficiency of Collisionless Accretion}. \apj
  501:787--+, \doi{10.1086/305845}, \eprint{arXiv:astro-ph/9710132}

\bibitem[{{G{\"u}ltekin} et~al(2009){G{\"u}ltekin}, {Richstone}, {Gebhardt},
  {Lauer}, {Tremaine}, {Aller}, {Bender}, {Dressler}, {Faber}, {Filippenko},
  {Green}, {Ho}, {Kormendy}, {Magorrian}, {Pinkney}, and
  {Siopis}}]{Gultekin2009}
{G{\"u}ltekin} K, {Richstone} DO, {Gebhardt} K, {Lauer} TR, {Tremaine} S,
  {Aller} MC, {Bender} R, {Dressler} A, {Faber} SM, {Filippenko} AV, {Green} R,
  {Ho} LC, {Kormendy} J, {Magorrian} J, {Pinkney} J, {Siopis} C (2009) {The
  M-{$\sigma$} and M-L Relations in Galactic Bulges, and Determinations of
  Their Intrinsic Scatter}. \apj 698:198--221,
  \doi{10.1088/0004-637X/698/1/198}, \eprint{0903.4897}

\bibitem[{{G{\"u}rkan} et~al(2004){G{\"u}rkan}, {Freitag}, and
  {Rasio}}]{ato2004}
{G{\"u}rkan} MA, {Freitag} M, {Rasio} FA (2004) {Formation of Massive Black
  Holes in Dense Star Clusters. I. Mass Segregation and Core Collapse}. ApJ
  604:632--652, \doi{10.1086/381968}, \eprint{astro-ph/0308449}

\bibitem[{{G{\"u}rkan} et~al(2006){G{\"u}rkan}, {Fregeau}, and
  {Rasio}}]{ato2006}
{G{\"u}rkan} MA, {Fregeau} JM, {Rasio} FA (2006) {Massive Black Hole Binaries
  from Collisional Runaways}. ApJL 640:L39--L42, \doi{10.1086/503295},
  \eprint{astro-ph/0512642}

\bibitem[{{Haehnelt} and {Rees}(1993)}]{haehnelt1993}
{Haehnelt} MG, {Rees} MJ (1993) {The formation of nuclei in newly formed
  galaxies and the evolution of the quasar population}. {MNRAS} 263:168--178

\bibitem[{{Haiman}(2004)}]{Haiman2004}
{Haiman} Z (2004) {Constraints from Gravitational Recoil on the Growth of
  Supermassive Black Holes at High Redshift}. {ApJ} 613:36--40,
  \doi{10.1086/422910}

\bibitem[{{H{\"a}ring} and {Rix}(2004)}]{Haring2004}
{H{\"a}ring} N, {Rix} HW (2004) {On the Black Hole Mass-Bulge Mass Relation}.
  \apjl 604:L89--L92, \doi{10.1086/383567}, \eprint{arXiv:astro-ph/0402376}

\bibitem[{{Hawking}(1971)}]{Hawking1971}
{Hawking} S (1971) {Gravitationally collapsed objects of very low mass}. \mnras
  152:75--+

\bibitem[{{Heger} et~al(2003){Heger}, {Fryer}, {Woosley}, {Langer}, and
  {Hartmann}}]{heger2003}
{Heger} A, {Fryer} CL, {Woosley} SE, {Langer} N, {Hartmann} DH (2003) {How
  Massive Single Stars End Their Life}. ApJ 591:288--300, \doi{10.1086/375341},
  \eprint{astro-ph/0212469}

\bibitem[{{Ho}(2008)}]{Ho2008}
{Ho} LC (2008) {Nuclear Activity in Nearby Galaxies}. \araa 46:475--539,
  \doi{10.1146/annurev.astro.45.051806.110546}, \eprint{0803.2268}

\bibitem[{{Hopkins} et~al(2007){Hopkins}, {Richards}, and
  {Hernquist}}]{Hopkins2007}
{Hopkins} PF, {Richards} GT, {Hernquist} L (2007) {An Observational
  Determination of the Bolometric Quasar Luminosity Function}. \apj
  654:731--753, \doi{10.1086/509629}, \eprint{arXiv:astro-ph/0605678}

\bibitem[{{Hoyle} and {Fowler}(1963)}]{hoyle1963}
{Hoyle} F, {Fowler} WA (1963) {On the nature of strong radio sources}. MNRAS
  125:169--+

\bibitem[{{Hughes}(2002)}]{Hughes2002}
{Hughes} SA (2002) {Untangling the merger history of massive black holes with
  LISA}. MNRAS 331:805--816, \doi{10.1046/j.1365-8711.2002.05247.x},
  \eprint{astro-ph/0108483}

\bibitem[{{Iocco}(2008)}]{Iocco2008}
{Iocco} F (2008) {Dark Matter Capture and Annihilation on the First Stars:
  Preliminary Estimates}. \apjl 677:L1--L4, \doi{10.1086/587959},
  \eprint{0802.0941}

\bibitem[{{Johnson} and {Bromm}(2007)}]{JBromm}
{Johnson} JL, {Bromm} V (2007) {The aftermath of the first stars: massive black
  holes}. \mnras 374:1557--1568, \doi{10.1111/j.1365-2966.2006.11275.x},
  \eprint{arXiv:astro-ph/0605691}

\bibitem[{{Khlopov} et~al(2005){Khlopov}, {Rubin}, and {Sakharov}}]{Russo2005}
{Khlopov} MY, {Rubin} SG, {Sakharov} AS (2005) {Primordial structure of massive
  black hole clusters}. Astroparticle Physics 23:265--277,
  \doi{10.1016/j.astropartphys.2004.12.002}, \eprint{arXiv:astro-ph/0401532}

\bibitem[{{Kormendy} et~al(1996){Kormendy}, {Bender}, {Richstone}, {Ajhar},
  {Dressler}, {Faber}, {Gebhardt}, {Grillmair}, {Lauer}, and
  {Tremaine}}]{Kormendy1996}
{Kormendy} J, {Bender} R, {Richstone} D, {Ajhar} EA, {Dressler} A, {Faber} SM,
  {Gebhardt} K, {Grillmair} C, {Lauer} TR, {Tremaine} S (1996) {Hubble Space
  Telescope Spectroscopic Evidence for a 2 billion solar masses Black Hole in
  NGC 3115}. \apjl 459:L57+, \doi{10.1086/309950}

\bibitem[{{Kormendy} et~al(2009){Kormendy}, {Fisher}, {Cornell}, and
  {Bender}}]{Kormendy2009}
{Kormendy} J, {Fisher} DB, {Cornell} ME, {Bender} R (2009) {Structure and
  Formation of Elliptical and Spheroidal Galaxies}. \apjs 182:216--309,
  \doi{10.1088/0067-0049/182/1/216}, \eprint{0810.1681}

\bibitem[{{Koushiappas} et~al(2004){Koushiappas}, {Bullock}, and
  {Dekel}}]{Koushiappas2004}
{Koushiappas} SM, {Bullock} JS, {Dekel} A (2004) {Massive black hole seeds from
  low angular momentum material}. {MNRAS} 354:292--304,
  \doi{10.1111/j.1365-2966.2004.08190.x}, \eprint{astro-ph/0311487}

\bibitem[{{Kudritzki} and {Puls}(2000)}]{Kudritzki2000}
{Kudritzki} RP, {Puls} J (2000) {Winds from Hot Stars}. \araa 38:613--666,
  \doi{10.1146/annurev.astro.38.1.613}

\bibitem[{{Lacey} and {Cole}(1993)}]{Lacey1993}
{Lacey} C, {Cole} S (1993) {Merger rates in hierarchical models of galaxy
  formation}. {MNRAS} 262:627--649

\bibitem[{{Lodato} and {Natarajan}(2006)}]{LN2006}
{Lodato} G, {Natarajan} P (2006) {Supermassive black hole formation during the
  assembly of pre-galactic discs}. {MNRAS} 371:1813--1823,
  \doi{10.1111/j.1365-2966.2006.10801.x}, \eprint{astro-ph/0606159}

\bibitem[{{Loeb} and {Rasio}(1994)}]{LoebRasio1994}
{Loeb} A, {Rasio} FA (1994) {Collapse of primordial gas clouds and the
  formation of quasar black holes}. {ApJ} 432:52--61, \doi{10.1086/174548},
  \eprint{astro-ph/9401026}

\bibitem[{{Madau} and {Rees}(2001)}]{MadauRees2001}
{Madau} P, {Rees} MJ (2001) {Massive Black Holes as Population III Remnants}.
  {ApJL} 551:L27--L30, \doi{10.1086/319848}

\bibitem[{{Magorrian} et~al(1998){Magorrian}, {Tremaine}, {Richstone},
  {Bender}, {Bower}, {Dressler}, {Faber}, {Gebhardt}, {Green}, {Grillmair},
  {Kormendy}, and {Lauer}}]{Magorrian1998}
{Magorrian} J, {Tremaine} S, {Richstone} D, {Bender} R, {Bower} G, {Dressler}
  A, {Faber} SM, {Gebhardt} K, {Green} R, {Grillmair} C, {Kormendy} J, {Lauer}
  T (1998) {The Demography of Massive Dark Objects in Galaxy Centers}. {AJ}
  115:2285--2305, \doi{10.1086/300353}

\bibitem[{{Majewski} et~al(2005){Majewski}, {Frinchaboy}, {Kunkel}, {Link},
  {Mu{\~n}oz}, {Ostheimer}, {Palma}, {Patterson}, and
  {Geisler}}]{Majewski:2005}
{Majewski} SR, {Frinchaboy} PM, {Kunkel} WE, {Link} R, {Mu{\~n}oz} RR,
  {Ostheimer} JC, {Palma} C, {Patterson} RJ, {Geisler} D (2005) {Exploring Halo
  Substructure with Giant Stars. VI. Extended Distributions of Giant Stars
  around the Carina Dwarf Spheroidal Galaxy: How Reliable Are They?} \aj
  130:2677--2700, \doi{10.1086/444535}, \eprint{arXiv:astro-ph/0503627}

\bibitem[{{Marconi} and {Hunt}(2003)}]{MarconiHunt2003}
{Marconi} A, {Hunt} LK (2003) {The Relation between Black Hole Mass, Bulge
  Mass, and Near-Infrared Luminosity}. \apjl 589:L21--L24,
  \doi{10.1086/375804}, \eprint{arXiv:astro-ph/0304274}

\bibitem[{{Marconi} et~al(2004){Marconi}, {Risaliti}, {Gilli}, {Hunt},
  {Maiolino}, and {Salvati}}]{Marconi2004}
{Marconi} A, {Risaliti} G, {Gilli} R, {Hunt} LK, {Maiolino} R, {Salvati} M
  (2004) {Local supermassive black holes, relics of active galactic nuclei and
  the X-ray background}. {MNRAS} 351:169--185,
  \doi{10.1111/j.1365-2966.2004.07765.x}

\bibitem[{Mateo(1998)}]{Mateo:1998}
Mateo M (1998) Dwarf galaxies of the local group. Annu Rev Astron Astrophys
  36:435--506

\bibitem[{{McKee} and {Tan}(2008)}]{McKee2008}
{McKee} CF, {Tan} JC (2008) {The Formation of the First Stars. II. Radiative
  Feedback Processes and Implications for the Initial Mass Function}. \apj
  681:771--797, \doi{10.1086/587434}, \eprint{0711.1377}

\bibitem[{{Merloni} and {Heinz}(2008)}]{Merloni08}
{Merloni} A, {Heinz} S (2008) {A synthesis model for AGN evolution:
  supermassive black holes growth and feedback modes}. \mnras 388:1011--1030,
  \doi{10.1111/j.1365-2966.2008.13472.x}, \eprint{0805.2499}

\bibitem[{{Merloni} et~al(2004){Merloni}, {Rudnick}, and {Di
  Matteo}}]{Merlonietal2004}
{Merloni} A, {Rudnick} G, {Di Matteo} T (2004) {Tracing the cosmological
  assembly of stars and supermassive black holes in galaxies}. \mnras
  354:L37--L42, \doi{10.1111/j.1365-2966.2004.08382.x}

\bibitem[{{Merritt} et~al(2001){Merritt}, {Ferrarese}, and
  {Joseph}}]{Merritt2001}
{Merritt} D, {Ferrarese} L, {Joseph} CL (2001) {No Supermassive Black Hole in
  M33?} Science 293:1116--1119, \doi{10.1126/science.1063896},
  \eprint{arXiv:astro-ph/0107359}

\bibitem[{{Miller} and {Hamilton}(2002)}]{Miller2002}
{Miller} MC, {Hamilton} DP (2002) {Production of intermediate-mass black holes
  in globular clusters}. \mnras 330:232--240,
  \doi{10.1046/j.1365-8711.2002.05112.x}, \eprint{arXiv:astro-ph/0106188}

\bibitem[{{Milosavljevi{\'c}} et~al(2009){Milosavljevi{\'c}}, {Couch}, and
  {Bromm}}]{Milos2009}
{Milosavljevi{\'c}} M, {Couch} SM, {Bromm} V (2009) {Accretion Onto
  Intermediate-Mass Black Holes in Dense Protogalactic Clouds}. \apjl
  696:L146--L149, \doi{10.1088/0004-637X/696/2/L146}, \eprint{0812.2516}

\bibitem[{{Mo} et~al(1998){Mo}, {Mao}, and {White}}]{MoMaoWhite1998}
{Mo} HJ, {Mao} S, {White} SDM (1998) {The formation of galactic discs}. {MNRAS}
  295:319--336

\bibitem[{{Narayan} et~al(2000){Narayan}, {Igumenshchev}, and
  {Abramowicz}}]{Narayan2000}
{Narayan} R, {Igumenshchev} IV, {Abramowicz} MA (2000) {Self-similar Accretion
  Flows with Convection}. \apj 539:798--808, \doi{10.1086/309268},
  \eprint{arXiv:astro-ph/9912449}

\bibitem[{{Navarro} et~al(1997){Navarro}, {Frenk}, and {White}}]{NFW1997}
{Navarro} JF, {Frenk} CS, {White} SDM (1997) {A Universal Density Profile from
  Hierarchical Clustering}. {ApJ} 490:493--+, \doi{10.1086/304888}

\bibitem[{{Oh} and {Haiman}(2002)}]{OhHaiman2002}
{Oh} SP, {Haiman} Z (2002) {Second-Generation Objects in the Universe:
  Radiative Cooling and Collapse of Halos with Virial Temperatures above
  $10^{4}$ K}. {ApJ} 569:558--572, \doi{10.1086/339393}

\bibitem[{{Omukai} and {Nishi}(1998)}]{Omukai1998}
{Omukai} K, {Nishi} R (1998) {Formation of Primordial Protostars}. \apj
  508:141--150, \doi{10.1086/306395}, \eprint{arXiv:astro-ph/9811308}

\bibitem[{{Omukai} et~al(2008){Omukai}, {Schneider}, and {Haiman}}]{omukai2008}
{Omukai} K, {Schneider} R, {Haiman} Z (2008) {Can Supermassive Black Holes Form
  in Metal-enriched High-Redshift Protogalaxies?} \apj 686:801--814,
  \doi{10.1086/591636}, \eprint{0804.3141}

\bibitem[{{Page} and {Hawking}(1976)}]{Page1976}
{Page} DN, {Hawking} SW (1976) {Gamma rays from primordial black holes}. \apj
  206:1--7, \doi{10.1086/154350}

\bibitem[{{Palla} et~al(2002){Palla}, {Zinnecker}, {Maeder}, and
  {Meynet}}]{palla2002}
{Palla} F, {Zinnecker} H, {Maeder} A, {Meynet} G (eds) (2002) {Physics of star
  formation in galaxies}

\bibitem[{{Peacock}(1999)}]{Peacock1999}
{Peacock} JA (1999) {Cosmological Physics}

\bibitem[{{Pelupessy} et~al(2007){Pelupessy}, {Di Matteo}, and
  {Ciardi}}]{pelupessy2007}
{Pelupessy} FI, {Di Matteo} T, {Ciardi} B (2007) {How rapidly do supermassive
  black hole ''seeds'' grow at early times?} ArXiv Astrophysics e-prints,
  astro-ph/0703773 \eprint{astro-ph/0703773}

\bibitem[{{Portegies Zwart} and {McMillan}(2002)}]{PZ2002}
{Portegies Zwart} SF, {McMillan} SLW (2002) {The Runaway Growth of
  Intermediate-Mass Black Holes in Dense Star Clusters}. ApJ 576:899--907,
  \doi{10.1086/341798}, \eprint{astro-ph/0201055}

\bibitem[{{Portegies Zwart} et~al(1999){Portegies Zwart}, {Makino}, {McMillan},
  and {Hut}}]{PZ1999}
{Portegies Zwart} SF, {Makino} J, {McMillan} SLW, {Hut} P (1999) {Star cluster
  ecology. III. Runaway collisions in young compact star clusters}. \aap
  348:117--126, \eprint{arXiv:astro-ph/9812006}

\bibitem[{{Portegies Zwart} et~al(2004){Portegies Zwart}, {Baumgardt}, {Hut},
  {Makino}, and {McMillan}}]{PZ2004}
{Portegies Zwart} SF, {Baumgardt} H, {Hut} P, {Makino} J, {McMillan} SLW (2004)
  {Formation of massive black holes through runaway collisions in dense young
  star clusters}. Nature 428:724--726, \doi{10.1038/nature02448},
  \eprint{astro-ph/0402622}

\bibitem[{{Quataert} and {Gruzinov}(2000)}]{Quataert2000}
{Quataert} E, {Gruzinov} A (2000) {Convection-dominated Accretion Flows}. \apj
  539:809--814, \doi{10.1086/309267}, \eprint{arXiv:astro-ph/9912440}

\bibitem[{{Reed} et~al(2005){Reed}, {Governato}, {Quinn}, {Gardner}, {Stadel},
  and {Lake}}]{Reed2005}
{Reed} D, {Governato} F, {Quinn} T, {Gardner} J, {Stadel} J, {Lake} G (2005)
  {Dark matter subhaloes in numerical simulations}. \mnras 359:1537--1548,
  \doi{10.1111/j.1365-2966.2005.09020.x}, \eprint{arXiv:astro-ph/0406034}

\bibitem[{{Rees}(1978)}]{Rees1978}
{Rees} MJ (1978) {Emission from the nuclei of nearby galaxies - Evidence for
  massive black holes}. In: {Berkhuijsen} EM, {Wielebinski} R (eds) Structure
  and Properties of Nearby Galaxies, IAU Symposium, vol~77, pp 237--242

\bibitem[{{Regan} and {Haehnelt}(2009)}]{Regan2009}
{Regan} JA, {Haehnelt} MG (2009) {Pathways to massive black holes and compact
  star clusters in pre-galactic dark matter haloes with virial temperatures
  >10000K}. \mnras 396:343--353, \doi{10.1111/j.1365-2966.2009.14579.x},
  \eprint{0810.2802}

\bibitem[{{Ricotti} et~al(2008){Ricotti}, {Ostriker}, and {Mack}}]{Ricotti2008}
{Ricotti} M, {Ostriker} JP, {Mack} KJ (2008) {Effect of Primordial Black Holes
  on the Cosmic Microwave Background and Cosmological Parameter Estimates}.
  \apj 680:829--845, \doi{10.1086/587831}, \eprint{0709.0524}

\bibitem[{{Ripamonti} et~al(2002){Ripamonti}, {Haardt}, {Ferrara}, and
  {Colpi}}]{Ripamonti2002}
{Ripamonti} E, {Haardt} F, {Ferrara} A, {Colpi} M (2002) {Radiation from the
  first forming stars}. \mnras 334:401--418,
  \doi{10.1046/j.1365-8711.2002.05516.x}, \eprint{arXiv:astro-ph/0107095}

\bibitem[{{Ripamonti} et~al(2007){Ripamonti}, {Mapelli}, and
  {Ferrara}}]{Ripamonti2007}
{Ripamonti} E, {Mapelli} M, {Ferrara} A (2007) {The impact of dark matter
  decays and annihilations on the formation of the first structures}. \mnras
  375:1399--1408, \doi{10.1111/j.1365-2966.2006.11402.x},
  \eprint{arXiv:astro-ph/0606483}

\bibitem[{{Saijo} et~al(2002){Saijo}, {Baumgarte}, {Shapiro}, and
  {Shibata}}]{saijo2002}
{Saijo} M, {Baumgarte} TW, {Shapiro} SL, {Shibata} M (2002) {Collapse of a
  Rotating Supermassive Star to a Supermassive Black Hole: Post-Newtonian
  Simulations}. ApJ 569:349--361, \doi{10.1086/339268},
  \eprint{astro-ph/0202112}

\bibitem[{{Salvaterra} et~al(2007){Salvaterra}, {Haardt}, and
  {Volonteri}}]{salvaterra07}
{Salvaterra} R, {Haardt} F, {Volonteri} M (2007) {Unresolved X-ray background:
  clues on galactic nuclear activity at z > 6}. MNRAS 374:761--768,
  \doi{10.1111/j.1365-2966.2006.11195.x}, \eprint{astro-ph/0610329}

\bibitem[{{Santoro} and {Shull}(2006)}]{santoro}
{Santoro} F, {Shull} JM (2006) {Critical Metallicity and Fine-Structure
  Emission of Primordial Gas Enriched by the First Stars}. ApJ 643:26--37,
  \doi{10.1086/501518}, \eprint{astro-ph/0509101}

\bibitem[{{Schneider} et~al(2006){Schneider}, {Omukai}, {Inoue}, and
  {Ferrara}}]{schneider2006}
{Schneider} R, {Omukai} K, {Inoue} AK, {Ferrara} A (2006) {Fragmentation of
  star-forming clouds enriched with the first dust}. MNRAS 369:1437--1444,
  \doi{10.1111/j.1365-2966.2006.10391.x}, \eprint{arXiv:astro-ph/0603766}

\bibitem[{{Sch{\"o}del} et~al(2003){Sch{\"o}del}, {Ott}, {Genzel}, {Eckart},
  {Mouawad}, and {Alexander}}]{Schodel2003}
{Sch{\"o}del} R, {Ott} T, {Genzel} R, {Eckart} A, {Mouawad} N, {Alexander} T
  (2003) {Stellar Dynamics in the Central Arcsecond of Our Galaxy}. \apj
  596:1015--1034, \doi{10.1086/378122}, \eprint{arXiv:astro-ph/0306214}

\bibitem[{{Sesana} et~al(2007){Sesana}, {Volonteri}, and {Haardt}}]{GW3}
{Sesana} A, {Volonteri} M, {Haardt} F (2007) {The imprint of massive black hole
  formation models on the LISA data stream}. MNRAS 377:1711--1716,
  \doi{10.1111/j.1365-2966.2007.11734.x}, \eprint{arXiv:astro-ph/0701556}

\bibitem[{{Shapiro}(2004)}]{Shapiro2004}
{Shapiro} SL (2004) {Formation of Supermassive Black Holes: Simulations in
  General Relativity}. In: {Ho} LC (ed) Coevolution of Black Holes and
  Galaxies, pp 103--+

\bibitem[{{Shapiro}(2005)}]{Shapiro2005}
{Shapiro} SL (2005) {Spin, Accretion, and the Cosmological Growth of
  Supermassive Black Holes}. {ApJ} 620:59--68, \doi{10.1086/427065}

\bibitem[{{Sheth} and {Tormen}(1999)}]{Sheth1999}
{Sheth} RK, {Tormen} G (1999) {Large-scale bias and the peak background split}.
  {MNRAS} 308:119--126

\bibitem[{{Shibata} and {Shapiro}(2002)}]{Shibata2002}
{Shibata} M, {Shapiro} SL (2002) {Collapse of a Rotating Supermassive Star to a
  Supermassive Black Hole: Fully Relativistic Simulations}. \apjl 572:L39--L43,
  \doi{10.1086/341516}, \eprint{arXiv:astro-ph/0205091}

\bibitem[{{Shlosman} et~al(1989){Shlosman}, {Frank}, and
  {Begelman}}]{Shlosman1989}
{Shlosman} I, {Frank} J, {Begelman} MC (1989) {Bars within bars - A mechanism
  for fuelling active galactic nuclei}. Nature 338:45--47,
  \doi{10.1038/338045a0}

\bibitem[{{Soltan}(1982)}]{Soltan1982}
{Soltan} A (1982) {Masses of quasars}. \mnras 200:115--122

\bibitem[{{Spaans} and {Silk}(2006)}]{Spaans2006}
{Spaans} M, {Silk} J (2006) {Pregalactic Black Hole Formation with an Atomic
  Hydrogen Equation of State}. \apj 652:902--906, \doi{10.1086/508444},
  \eprint{arXiv:astro-ph/0601714}

\bibitem[{{Spitzer}(1987)}]{Spitzer1987}
{Spitzer} L (1987) {Dynamical evolution of globular clusters}

\bibitem[{{Spolyar} et~al(2008){Spolyar}, {Freese}, and
  {Gondolo}}]{Spolyar2008}
{Spolyar} D, {Freese} K, {Gondolo} P (2008) {Dark Matter and the First Stars: A
  New Phase of Stellar Evolution}. Physical Review Letters 100(5):051,101--+,
  \doi{10.1103/PhysRevLett.100.051101}, \eprint{0705.0521}

\bibitem[{{Springel} et~al(2008){Springel}, {Wang}, {Vogelsberger}, {Ludlow},
  {Jenkins}, {Helmi}, {Navarro}, {Frenk}, and {White}}]{Springel2008}
{Springel} V, {Wang} J, {Vogelsberger} M, {Ludlow} A, {Jenkins} A, {Helmi} A,
  {Navarro} JF, {Frenk} CS, {White} SDM (2008) {The Aquarius Project: the
  subhaloes of galactic haloes}. \mnras 391:1685--1711,
  \doi{10.1111/j.1365-2966.2008.14066.x}, \eprint{0809.0898}

\bibitem[{{Stacy} et~al(2009){Stacy}, {Greif}, and {Bromm}}]{Stacy2009}
{Stacy} A, {Greif} TH, {Bromm} V (2009) {The first stars: formation of binaries
  and small multiple systems}. ArXiv e-prints \eprint{0908.0712}

\bibitem[{{Tan} and {McKee}(2004)}]{Tan2004}
{Tan} JC, {McKee} CF (2004) {The Formation of the First Stars. I. Mass Infall
  Rates, Accretion Disk Structure, and Protostellar Evolution}. \apj
  603:383--400, \doi{10.1086/381490}, \eprint{arXiv:astro-ph/0307414}

\bibitem[{{Tanaka} and {Haiman}(2009)}]{Tanaka2009}
{Tanaka} T, {Haiman} Z (2009) {The Assembly of Supermassive Black Holes at High
  Redshifts}. \apj 696:1798--1822, \doi{10.1088/0004-637X/696/2/1798},
  \eprint{0807.4702}

\bibitem[{{Tegmark} et~al(1997){Tegmark}, {Silk}, {Rees}, {Blanchard}, {Abel},
  and {Palla}}]{Tegmark1997}
{Tegmark} M, {Silk} J, {Rees} MJ, {Blanchard} A, {Abel} T, {Palla} F (1997)
  {How Small Were the First Cosmological Objects?} \apj 474:1--+,
  \doi{10.1086/303434}, \eprint{arXiv:astro-ph/9603007}

\bibitem[{{Tisserand} et~al(2007){Tisserand}, {Le Guillou}, {Afonso}, {Albert},
  {Andersen}, {Ansari}, {Aubourg}, {Bareyre}, {Beaulieu}, {Charlot},
  {Coutures}, {Ferlet}, {Fouqu{\'e}}, {Glicenstein}, {Goldman}, {Gould},
  {Graff}, {Gros}, {Haissinski}, {Hamadache}, {de Kat}, {Lasserre}, {Lesquoy},
  {Loup}, {Magneville}, {Marquette}, {Maurice}, {Maury}, {Milsztajn}, {Moniez},
  {Palanque-Delabrouille}, {Perdereau}, {Rahal}, {Rich}, {Spiro},
  {Vidal-Madjar}, {Vigroux}, {Zylberajch}, and {The EROS-2
  Collaboration}}]{Tisserand2007}
{Tisserand} P, {Le Guillou} L, {Afonso} C, {Albert} JN, {Andersen} J, {Ansari}
  R, {Aubourg} {\'E}, {Bareyre} P, {Beaulieu} JP, {Charlot} X, {Coutures} C,
  {Ferlet} R, {Fouqu{\'e}} P, {Glicenstein} JF, {Goldman} B, {Gould} A, {Graff}
  D, {Gros} M, {Haissinski} J, {Hamadache} C, {de Kat} J, {Lasserre} T,
  {Lesquoy} {\'E}, {Loup} C, {Magneville} C, {Marquette} JB, {Maurice} {\'E},
  {Maury} A, {Milsztajn} A, {Moniez} M, {Palanque-Delabrouille} N, {Perdereau}
  O, {Rahal} YR, {Rich} J, {Spiro} M, {Vidal-Madjar} A, {Vigroux} L,
  {Zylberajch} S, {The EROS-2 Collaboration} (2007) {Limits on the Macho
  content of the Galactic Halo from the EROS-2 Survey of the Magellanic
  Clouds}. \aap 469:387--404, \doi{10.1051/0004-6361:20066017},
  \eprint{arXiv:astro-ph/0607207}

\bibitem[{{Tremaine} et~al(2002){Tremaine}, {Gebhardt}, {Bender}, {Bower},
  {Dressler}, {Faber}, {Filippenko}, {Green}, {Grillmair}, {Ho}, {Kormendy},
  {Lauer}, {Magorrian}, {Pinkney}, and {Richstone}}]{Tremaineetal2002}
{Tremaine} S, {Gebhardt} K, {Bender} R, {Bower} G, {Dressler} A, {Faber} SM,
  {Filippenko} AV, {Green} R, {Grillmair} C, {Ho} LC, {Kormendy} J, {Lauer} TR,
  {Magorrian} J, {Pinkney} J, {Richstone} D (2002) {The Slope of the Black Hole
  Mass versus Velocity Dispersion Correlation}. {ApJ} 574:740--753,
  \doi{10.1086/341002}

\bibitem[{{Trenti} et~al(2009){Trenti}, {Stiavelli}, and {Michael
  Shull}}]{Trenti2009}
{Trenti} M, {Stiavelli} M, {Michael Shull} J (2009) {Metal-free Gas Supply at
  the Edge of Reionization: Late-epoch Population III Star Formation}. \apj
  700:1672--1679, \doi{10.1088/0004-637X/700/2/1672}, \eprint{0905.4504}

\bibitem[{{Turk} et~al(2009){Turk}, {Abel}, and {O'Shea}}]{Turk2009}
{Turk} MJ, {Abel} T, {O'Shea} B (2009) {The Formation of Population III
  Binaries from Cosmological Initial Conditions}. Science 325:601--,
  \doi{10.1126/science.1173540}, \eprint{0907.2919}

\bibitem[{{Valluri} et~al(2005){Valluri}, {Ferrarese}, {Merritt}, and
  {Joseph}}]{Valluri2005}
{Valluri} M, {Ferrarese} L, {Merritt} D, {Joseph} CL (2005) {The Low End of the
  Supermassive Black Hole Mass Function: Constraining the Mass of a Nuclear
  Black Hole in NGC 205 via Stellar Kinematics}. \apj 628:137--152,
  \doi{10.1086/430752}, \eprint{arXiv:astro-ph/0502493}

\bibitem[{{van den Bosch} et~al(2002){van den Bosch}, {Abel}, {Croft},
  {Hernquist}, and {White}}]{VandenBosch2002}
{van den Bosch} FC, {Abel} T, {Croft} RAC, {Hernquist} L, {White} SDM (2002)
  {The Angular Momentum of Gas in Protogalaxies. I. Implications for the
  Formation of Disk Galaxies}. ApJ 576:21--35, \doi{10.1086/341619},
  \eprint{astro-ph/0201095}

\bibitem[{{Van Wassenhove} et~al(2010){Van Wassenhove}, {Volonteri}, {Walker},
  and {Gair}}]{Svanwas2010}
{Van Wassenhove} S, {Volonteri} M, {Walker} MG, {Gair} JR (2010) {Massive black
  holes lurking in Milky Way satellites}. ArXiv e-prints \eprint{1001.5451}

\bibitem[{{Vernaleo} and {Reynolds}(2006)}]{Vernaleo2006}
{Vernaleo} JC, {Reynolds} CS (2006) {AGN Feedback and Cooling Flows: Problems
  with Simple Hydrodynamic Models}. \apj 645:83--94, \doi{10.1086/504029},
  \eprint{arXiv:astro-ph/0511501}

\bibitem[{{Volonteri} and {Gnedin}(2009)}]{VG2009}
{Volonteri} M, {Gnedin} N (2009) {Relative Role of Stars and Quasars in Cosmic
  Reionization}. ArXiv e-prints \eprint{0905.0144}

\bibitem[{{Volonteri} and {Natarajan}(2009)}]{VN09}
{Volonteri} M, {Natarajan} P (2009) {Journey to the M\_BH -sigma relation: the
  fate of low mass black holes in the Universe}. ArXiv e-prints
  \eprint{0903.2262}

\bibitem[{{Volonteri} and {Rees}(2005)}]{VolonteriRees2005}
{Volonteri} M, {Rees} MJ (2005) {Rapid Growth of High-Redshift Black Holes}.
  {ApJ} 633:624--629, \doi{10.1086/466521}

\bibitem[{{Volonteri} and {Rees}(2006)}]{VRees2006}
{Volonteri} M, {Rees} MJ (2006) {Quasars at z=6: The Survival of the Fittest}.
  ApJ 650:669--678, \doi{10.1086/507444}, \eprint{arXiv:astro-ph/0607093}

\bibitem[{{Volonteri} et~al(2003){Volonteri}, {Haardt}, and {Madau}}]{VHM}
{Volonteri} M, {Haardt} F, {Madau} P (2003) {The Assembly and Merging History
  of Supermassive Black Holes in Hierarchical Models of Galaxy Formation}.
  {ApJ} 582:559--573, \doi{10.1086/344675}

\bibitem[{{Volonteri} et~al(2008{\natexlab{a}}){Volonteri}, {Haardt}, and
  {G{\"u}ltekin}}]{Volonterietal08}
{Volonteri} M, {Haardt} F, {G{\"u}ltekin} K (2008{\natexlab{a}}) {Compact
  massive objects in Virgo galaxies: the black hole population}. MNRAS
  384:1387--1392, \doi{10.1111/j.1365-2966.2008.12911.x},
  \eprint{arXiv:0710.5770}

\bibitem[{{Volonteri} et~al(2008{\natexlab{b}}){Volonteri}, {Lodato}, and
  {Natarajan}}]{VLN2008}
{Volonteri} M, {Lodato} G, {Natarajan} P (2008{\natexlab{b}}) {The evolution of
  massive black hole seeds}. MNRAS 383:1079--1088,
  \doi{10.1111/j.1365-2966.2007.12589.x}, \eprint{arXiv:0709.0529}

\bibitem[{{Walker} et~al(2009{\natexlab{a}}){Walker}, {Mateo}, and
  {Olszewski}}]{walker09a}
{Walker} MG, {Mateo} M, {Olszewski} EW (2009{\natexlab{a}}) {Stellar Velocities
  in the Carina, Fornax, Sculptor, and Sextans dSph Galaxies: Data From the
  Magellan/MMFS Survey}. \aj 137:3100--3108,
  \doi{10.1088/0004-6256/137/2/3100}, \eprint{0811.0118}

\bibitem[{{Walker} et~al(2009{\natexlab{b}}){Walker}, {Mateo}, and
  {Olszewski}}]{Walker2009}
{Walker} MG, {Mateo} M, {Olszewski} EW (2009{\natexlab{b}}) {Stellar Velocities
  in the Carina, Fornax, Sculptor, and Sextans dSph Galaxies: Data From the
  Magellan/MMFS Survey}. \aj 137:3100--3108,
  \doi{10.1088/0004-6256/137/2/3100}, \eprint{0811.0118}

\bibitem[{{Walter} et~al(2004){Walter}, {Carilli}, {Bertoldi}, {Menten}, {Cox},
  {Lo}, {Fan}, and {Strauss}}]{Walteretal2004}
{Walter} F, {Carilli} C, {Bertoldi} F, {Menten} K, {Cox} P, {Lo} KY, {Fan} X,
  {Strauss} MA (2004) {Resolved Molecular Gas in a Quasar Host Galaxy at
  Redshift z=6.42}. {ApJL} 615:L17--L20, \doi{10.1086/426017}

\bibitem[{{Wehner} and {Harris}(2006)}]{Wehner2006}
{Wehner} EH, {Harris} WE (2006) {From Supermassive Black Holes to Dwarf
  Elliptical Nuclei: A Mass Continuum}. {ApJL} 644:L17--L20,
  \doi{10.1086/505387}, \eprint{astro-ph/0603801}

\bibitem[{{White} and {Rees}(1978)}]{White1978}
{White} SDM, {Rees} MJ (1978) {Core condensation in heavy halos - A two-stage
  theory for galaxy formation and clustering}. \mnras 183:341--358

\bibitem[{{Willott} et~al(2003){Willott}, {McLure}, and
  {Jarvis}}]{Willottetal2003}
{Willott} CJ, {McLure} RJ, {Jarvis} MJ (2003) {A 3 Billion Solar Masses Black
  Hole in the Quasar SDSS J1148+5251 at z=6.41}. \apjl 587:L15--L18,
  \doi{10.1086/375126}

\bibitem[{{Willott} et~al(2005){Willott}, {Percival}, {McLure}, {Crampton},
  {Hutchings}, {Jarvis}, {Sawicki}, and {Simard}}]{Willottetal2005}
{Willott} CJ, {Percival} WJ, {McLure} RJ, {Crampton} D, {Hutchings} JB,
  {Jarvis} MJ, {Sawicki} M, {Simard} L (2005) {Imaging of SDSS z>6 Quasar
  Fields: Gravitational Lensing, Companion Galaxies, and the Host Dark Matter
  Halos}. {ApJ} 626:657--665, \doi{10.1086/430168}

\bibitem[{{Wise} et~al(2008){Wise}, {Turk}, and {Abel}}]{Wise2008}
{Wise} JH, {Turk} MJ, {Abel} T (2008) {Resolving the Formation of
  Protogalaxies. II. Central Gravitational Collapse}. \apj 682:745--757,
  \doi{10.1086/588209}, \eprint{0710.1678}

\bibitem[{{Woosley} and {Weaver}(1986)}]{Woosley1986}
{Woosley} SE, {Weaver} TA (1986) {The physics of supernova explosions}. \araa
  24:205--253, \doi{10.1146/annurev.aa.24.090186.001225}

\bibitem[{{Yoshida} et~al(2006){Yoshida}, {Omukai}, {Hernquist}, and
  {Abel}}]{Yoshida2006}
{Yoshida} N, {Omukai} K, {Hernquist} L, {Abel} T (2006) {Formation of
  Primordial Stars in a {$\Lambda$}CDM Universe}. \apj 652:6--25,
  \doi{10.1086/507978}, \eprint{arXiv:astro-ph/0606106}

\bibitem[{{Yu} and {Tremaine}(2002)}]{YuTremaine2002}
{Yu} Q, {Tremaine} S (2002) {Observational constraints on growth of massive
  black holes}. {MNRAS} 335:965--976, \doi{10.1046/j.1365-8711.2002.05532.x}

\bibitem[{{Zel'Dovich} and {Novikov}(1967)}]{Zeldovich1967}
{Zel'Dovich} YB, {Novikov} ID (1967) {The Hypothesis of Cores Retarded during
  Expansion and the Hot Cosmological Model}. Soviet Astronomy 10:602--+

\bibitem[{{Zeldovich} and {Novikov}(1971)}]{Zeldovich1971}
{Zeldovich} YB, {Novikov} ID (1971) {Relativistic astrophysics. Vol.1: Stars
  and relativity}

\bibitem[{{Zhang} et~al(2008){Zhang}, {Woosley}, and {Heger}}]{Zhang2008}
{Zhang} W, {Woosley} SE, {Heger} A (2008) {Fallback and Black Hole Production
  in Massive Stars}. \apj 679:639--654, \doi{10.1086/526404},
  \eprint{arXiv:astro-ph/0701083}

\end{thebibliography}
\end{document}